\documentclass[11pt]{article}
\usepackage{amsmath}
\usepackage{amsfonts}
\usepackage{amssymb}
\usepackage{graphicx}
\usepackage{enumerate}
\usepackage{natbib}
\usepackage{url} 
\usepackage{multirow}
\usepackage{booktabs}

\usepackage[version=4]{mhchem}
\usepackage{siunitx}
\usepackage{bm}
\usepackage{graphicx}
\usepackage{caption}
\usepackage{subcaption}
\usepackage{lineno}
\usepackage{lipsum}

\newcommand\blfootnote[1]{%
  \begingroup
  \renewcommand\thefootnote{}\footnote{#1}%
  \addtocounter{footnote}{-1}%
  \endgroup
}

\newcommand{\blind}{0}

\begin{document}

\bibliographystyle{agsm}

\def\spacingset#1{\renewcommand{\baselinestretch}%
{#1}\small\normalsize} \spacingset{1}


\if0\blind
{
  \title{\bf A Mechanistic Model of Annual Sulfate Concentrations in the United States} 
  \author{Nathan B. Wikle$^1$\thanks{Correspondence to: N.B. Wikle, Email: nbwikle@psu.edu}, Ephraim M. Hanks$^1$, Lucas R.F. Henneman$^2$, and Corwin M. Zigler$^3$}
\date{%
    $^1$Department of Statistics, Pennsylvania State University\\%
    $^2$Department of Civil, Environ., and Infrastructure Engineering, George Mason University\\%
    $^3$Department of Statistics and Data Sciences, University of Texas at Austin\\[2ex]%
    \today
}
  \maketitle
} \fi

\if1\blind
{
{
\title{\bf A Mechanistic Model of Annual Sulfate Concentrations in the United States}
}
  \maketitle
} \fi

\spacingset{1}

\begin{abstract}
We develop a mechanistic model to analyze the impact of sulfur dioxide emissions from coal-fired power plants on average sulfate concentrations in the central United States.  A multivariate Ornstein-Uhlenbeck (OU) process is used to approximate the dynamics of the underlying space-time chemical transport process, and its distributional properties are leveraged to specify novel probability models for spatial data (i.e., spatially-referenced data with no temporal replication) that are viewed as either a snapshot or a time-averaged observation of the OU process. Air pollution transport dynamics determine the mean and covariance structure of our atmospheric sulfate model, allowing us to infer which process dynamics are driving observed air pollution concentrations. We use these inferred dynamics to assess the regulatory impact of flue-gas desulfurization (FGD) technologies on human exposure to sulfate aerosols. 
\end{abstract}

\noindent%


\section{Introduction}
\label{sec:intro}

Sulfur\blfootnote{Supplemental material available at \url{https://github.com/nbwikle/mechanisticSO4-supp_material}} dioxide (\ce{SO2}) emissions from coal-fired power plants are a major source of anthropogenic air pollution \citep{Rowe1980}.  Upon release into the atmosphere, \ce{SO2} compounds are simultaneously acted upon by chemical and physical processes, forming particulate sulfates (\ce{SO4^{2-}}) which are then transported across space.  Exposure to sulfate aerosols is associated with many adverse human health outcomes, including decreased lung function \citep{Ng2019}, increased risk of cardiovascular disease \citep{Bai2019heart}, and lung cancer \citep{Bai2019cancer}. Sulfate also represents one component of fine particulate matter, regulations for which comprise the most beneficial -- and costly -- monetized federal regulations in the United States \citep{Dominici2014}. Consequently, assessing the regulatory impacts of various \ce{SO2} emissions scenarios on ambient \ce{SO4^{2-}} concentrations may lead to an improved understanding of regional public health outcomes. 

Importantly, the dependence of air pollution concentrations on upwind emissions sources suggests that known atmospheric pollution transport dynamics should be included in the analysis of spatial air pollution data. The inclusion of process dynamics is especially important when inferring the effect of power plant emissions on observed public health outcomes across a large spatial domain \citep{ZiglerPapadogeorgou2018}. Existing atmospheric pollution analyses often fall into two camps: numerical models, such as chemical transport models \citep{Stein2015}, plume models, and their so-called reduced-form hybrids \citep{Foley2014, Heo2016}, in which process dynamics are used to simulate many individual-level trajectories of point-source pollutants, and phenomenological statistical models which seek to accurately interpolate regional air pollution concentrations from a variety of monitoring systems \citep{vanDonkelaar2019, Guan2020}. Although these models have proved useful, their utility is limited by the lack of data-driven inference in the numerical models, and the lack of known mechanistic processes in the statistical models, respectively. In this paper, we seek to develop a statistical model of yearly-aggregated \ce{SO4^2-} concentrations attributed to coal-fired power plant \ce{SO2} emissions, in which the known physical processes governing pollution transport directly inform the first and second-order structure of our model. The advantages of this modeling approach include its ability to infer process dynamics and stochastic fluctuations on a time-scale of interest and provide reasonable forecasts of future air pollution levels from spatial data alone. In doing so, we construct a new class of \textit{mechanistic spatial models} which allows us to model spatial \ce{SO4^2-} measurements based on a spatio-temporal generating model appropriate for this system.

This class of models is also beneficial when modeling spatial data from a variety of scientific systems, as most spatial data are best understood as arising from a spatio-temporal generating process.  For example, spatial genetic data are observed at a specific point in time, but the evolutionary process, driven by gene flow and migration, is inherently spatio-temporal \citep{Manel2003, Hanks2017}.  Similarly, the spatial distribution of a viral infectious disease, influenced by past vaccination and demographic patterns, is only a single temporal snapshot of a dynamic epidemic process  \citep{KeelingRohani2008}.   When spatial observations are made repeatedly in time, science-based dynamic spatio-temporal models have been used to great effect \citep{WikleHooten2010, CressieWikle2011}, often enabling more accurate probabilistic forecasts and scientific insights than traditional descriptive models \citep{Hefley2017}.

In contrast, statistical models for explicitly spatial data -- i.e., spatially-referenced data observed without temporal replication -- are often chosen without consideration of the underlying dynamic process.  Instead, spatial statistical models are mostly phenomenological, focused on modeling (1) the correlation between the process mean and (local) covariates, and (2) residual autocorrelation through the addition of a spatial random effect \citep{Banerjee2004}.   The random effect is often chosen to have a Gaussian process (GP) prior distribution, where the GP covariance is dictated by the spatial support of the data.  Common choices include the Matern class of covariance functions for point-referenced spatial data (cf. \citet{Cressie1993}), and conditional autoregressive (CAR) \citep{Besag1974, BesagKooperberg1994, Cressie1993} or simultaneous autoregressive (SAR) \citep{VerHoef2018, Cressie1993} covariance structures for areal/lattice data.  Although some classes of GP covariance functions have links to differential equations (see, e.g., \citet{Whittle1954, Whittle1962} and \citet{Lindgren2011}), in practice the predominant consideration when choosing the covariance of the spatial random effect remains the (spatial) support of the data.  This practice is reasonable when the spatial data are generated from a process with unknown or nonexistent temporal dynamics.  However, in many scientific systems it is likely that both the mean structure and observed pattern of spatial autocorrelation in the spatial data are influenced by a known space-time generating process.  

Consequently, the analysis of spatial data from processes with well-understood dynamics stands to benefit from more \textit{mechanistic} modeling approaches, in which knowledge of the process dynamics implies a natural likelihood model for the spatial data.  \citet{Hanks2017} suggests one approach for constructing mechanistic spatial models: consider a science-based spatio-temporal process and find its stationary distribution.  In particular, \citet{Hanks2017} constructs a spatial model for genetic allele data by deriving the stationary distribution of an asymmetric random walk model for gene flow.  The stationary distribution is a random field with discrete spatial support and a SAR covariance structure defined by the (parameterized) random walk model.  The resulting mechanistic model was able to infer spatio-temporal movement rates from spatial data alone, providing important scientific insight that is unobtainable when using a standard semiparametric random effect model.  While this approach opens the door to building spatial models that directly result from spatio-temporal processes, the processes considered were limited. In particular, they were not allowed to have time-varying noise; the spatial process was assumed to have reached a stationary distribution, at which it would continue constant in time.

In this paper, we construct a general class of mechanistic spatial models from the well-studied Ornstein-Uhlenbeck (OU) process.  These mechanistic models can accommodate spatial data viewed as either a transient or stationary `snapshot' of a spatio-temporal process, or as a time-averaged observation of a space-time process over a finite time interval.  Many common spatial data can be seen as a special case of one of these scenarios.  The models we develop are flexible enough to handle many linear dynamic systems, and space-time noise is explicitly included in the OU process construction.   Importantly, both the process dynamics and temporal window of observation dictate the mean and spatial covariance structure of the model; in special cases, these resemble familiar specifications of spatial autocorrelation. Our development of these methods is motivated by an analysis of the impact of coal-fired power plant sulfur dioxide (\ce{SO2}) emissions on average atmospheric sulfate (\ce{SO4^2-}) concentrations across the central U.S.; the mechanistic spatial model allows for probabilistic forecasts of average \ce{SO4^2-} concentrations under alternative emissions scenarios that are unavailable when using traditional phenomenological models. Such statistical forecasting with complete uncertainty quantification promises to advance current methodology for environmental risk assessment that relies on deterministic physical-chemical models to project pollution under various counterfactual scenarios. 

The remainder of the article is organized as follows.  In Section~\ref{sec:data}, we provide an overview of the particulate sulfate data and its connection to coal-fired power plant emissions. We then present a dynamic model of atmospheric sulfate concentration as a motivating space-time process from which spatial data are observed.  In Section~\ref{sec:spacetoOU}, we construct a general framework for analyzing spatial data from a broad class of space-time processes, emphasizing how process dynamics and the type of spatial observation can be leveraged to specify mechanistic models for spatial data.  We then analyze the 2011 average sulfate pollution in the central U.S. attributed to sulfur dioxide emissions from power plants (Section~\ref{sec:analysis}).  Finally, in Section~\ref{sec:disc} we discuss possible extensions of our mechanistic approach for modeling spatial air pollution data.


\section{Atmospheric Sulfate Concentration from Power Plant Emissions}
\label{sec:data}

Given the known association between exposure to fine particulate matter (including sulfate) and a number of chronic diseases \citep{Brook2010}, the United States has implemented monetized federal regulations to limit emissions from sources of anthropogenic air pollution \citep{Dominici2014}. Consequently, estimating the effect of such regulation on regional public health outcomes is a major effort in environmental epidemiology. However, because air pollution concentrations may be heavily dependent on upwind emissions source, it is often critical that the (nonstationary) dependence between pollution sources and concentrations be accounted for in such an analysis. For example, \citet{ZiglerPapadogeorgou2018} estimate the causal effect of reduced nitrous oxides emissions on hospitalization rates for cardiovascular disease across the United States.  To account for possible nitrous oxide exposure from multiple upwind power plants, they develop a framework for bipartite causal inference with interference, where treatments are defined on one type of interventional unit (e.g., power plants) and outcomes are defined for separate observational units, such as regional health outcomes.  \citet{ZiglerPapadogeorgou2018} incorporate the pollution transport dynamics through their definition of spatial interference, where potential outcomes for a particular region (hospitalization rates) are dependent upon the treatments (emissions reductions) assigned to multiple upwind power plants. In other words, the emissions from a single power plant propagate across space, impacting health outcomes across a large spatial extent.  Without accounting for this complex dependence structure, statistical inference on the health or environmental impacts of power plant emissions is limited.

\begin{figure}[ht]
\centering
    \begin{subfigure}[b]{0.4\textwidth}
        \includegraphics[width=\textwidth]{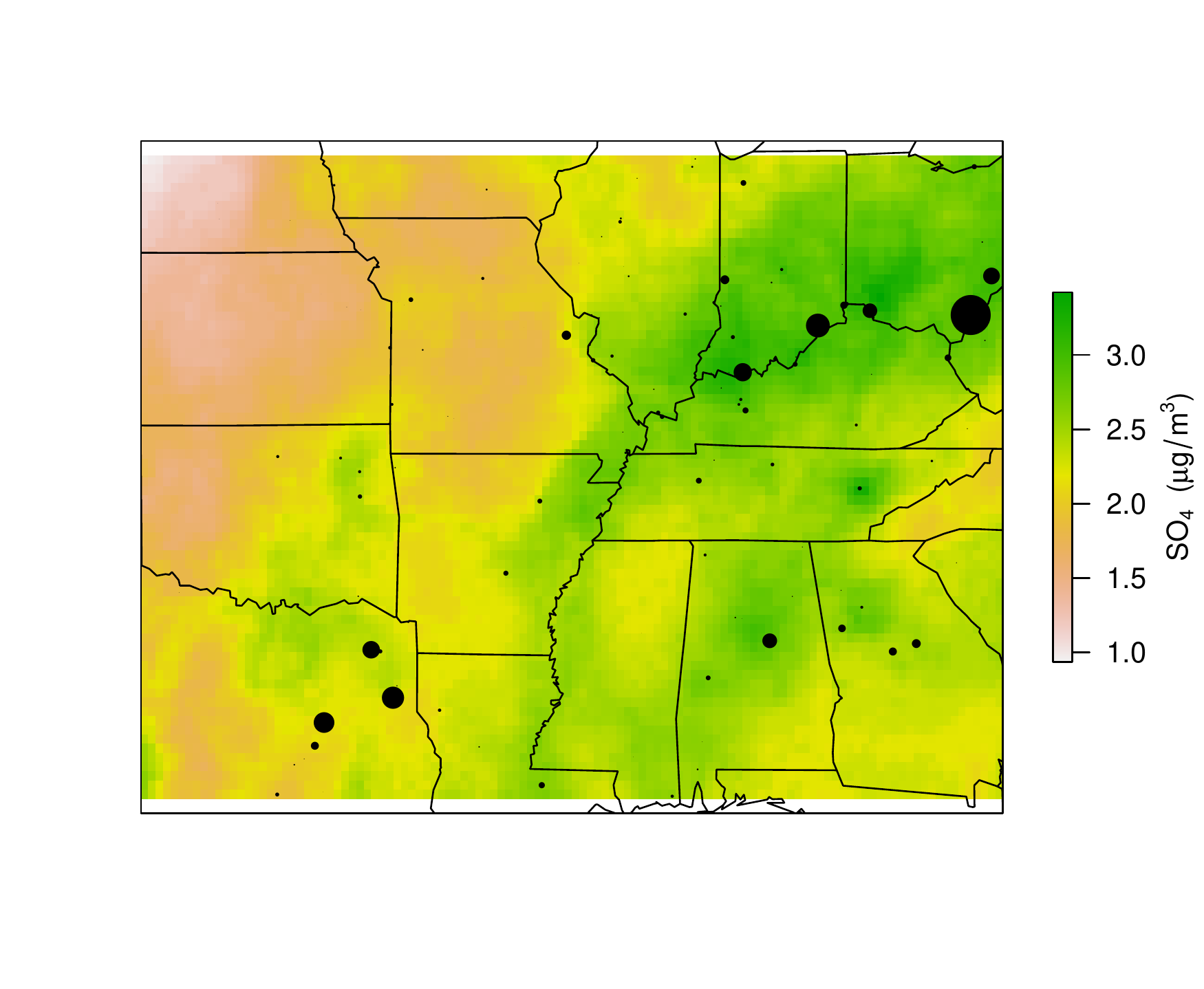}
        \caption{\ce{SO2} emissions sources.}
        \label{subfig:emissions}
    \end{subfigure}
    ~ 
    \begin{subfigure}[b]{0.4\textwidth}
        \includegraphics[width=\textwidth]{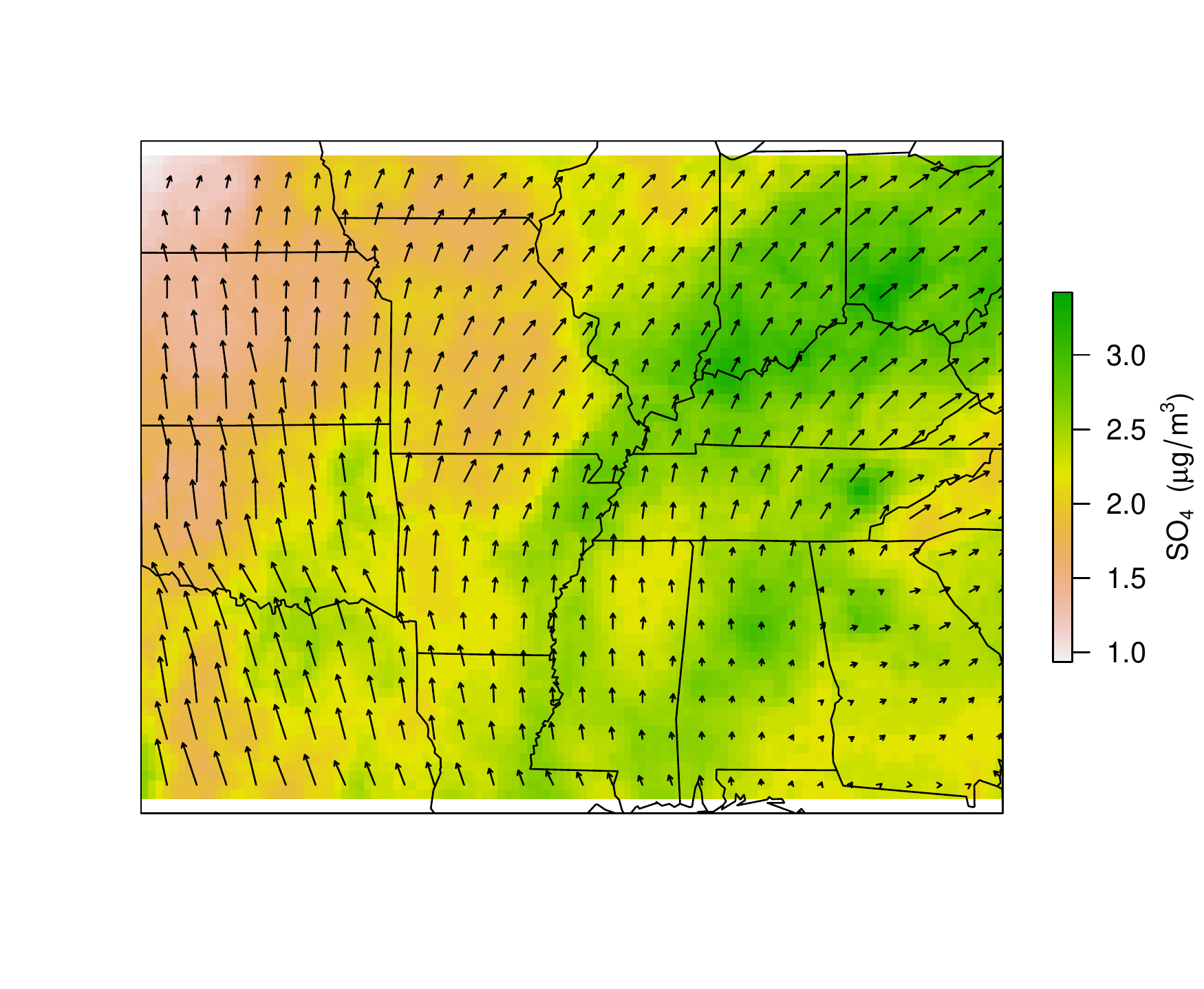}
        \caption{Average yearly wind velocities (10 \si{m}).}
        \label{subfig:wind}
    \end{subfigure}
    \caption{Average 2011 sulfate concentrations (\si{\ug / m^3}) across the central U.S., including (a) coal-fired power plant locations, weighted by total \ce{SO2} emissions, and (b) the 2011 average wind velocities (height = 10 \si{m}).}
    \label{fig:2011data}
\end{figure}

With this spatial interference in mind, we analyzed average 2011 sulfate concentrations across the central U.S. (Figure~\ref{fig:2011data}).   Estimated annual average sulfate concentrations (\si{\ug / m^3}) were obtained from the Dalhousie University Atmospheric Composition Analysis Group, version V4.NA.02 \citep{vanDonkelaar2019}.  These data were supplemented with annual coal-fired power plant \ce{SO2} emissions totals from the U.S. EPA's Air Markets Program Database (AMPD) \citep{EPA2016, Henneman2019}, as well as with meteorological data (e.g., average wind velocity, precipitation, etc.) from the NCEP/NCAR reanalysis database \citep{Kalnay1996}. In our analysis, we assess the influence of annual sulfur dioxide emissions totals (Figure~\ref{subfig:emissions}) and average wind velocity (Figure~\ref{subfig:wind}) on 2011 sulfate levels.  We propose a mechanistic modeling approach to account for the potentially large spatial footprint of upwind emissions sources.  In addition, we provide probabilistic forecasts of \ce{SO4^2-} concentrations under alternative emissions scenarios, and identify the regulatory policy which is expected to best reduce human exposure to atmospheric \ce{SO4^2-}.  In order to construct a mechanistic model for spatial air pollution data, we first present a model for the space-time emissions process.
  
\subsection{A Space-Time Model for Sulfate Concentration}
\label{sec:ctm}
 
The physical dynamics governing sulfate aerosols can be approximated with a mechanistic representation of atmospheric chemical transport.  Note that the development of realistic chemical transport models is a broad and active area of research in atmospheric chemistry (see, e.g., \citet{SeinfeldPandis2016} for details).  However, for the purposes of this analysis, we consider a relatively simple model of sulfate transport that necessarily sacrifices complexity for increased spatial and temporal scaleability, paralleling recent trends in atmospheric modeling using so-called reduced-complexity models \citep{Tessum2017}. The model characterizes changes in aerosol concentration as a result of four key processes: advection-diffusion, atmospheric deposition (dry and wet), chemical reaction, and emission.  These processes can be specified with a differential equation augmented by a mean-zero white noise process; the addition of uncertainty (e.g., to account for shifting weather patterns, landscape variability, unaccounted \ce{SO4^2-} sources, etc.) results in a stochastic partial differential equation (SPDE).
 
Let $y_i(t) \equiv y(\bm{s}_i, t)$ denote the atmospheric \ce{SO4^2-} concentration at location $\bm{s}_i \in \mathcal{D}$ and time $t$, where $\mathcal{D} \subset \mathbb{R}^2$ is the continuous spatial domain of interest.  Formally, the governing equation of the process can be written as
\begin{equation}
 d y_i(t) = \bigg[ \big( \gamma \Delta + \alpha \; \bm{v}_i(t) \cdot \nabla - \delta \big) y_i(t) + S_i(t) \bigg] dt + \sigma \, \xi(\bm{s}, t),\label{eq:transportmodel}
\end{equation}
where $\gamma \Delta$ denotes homogeneous spatial diffusion with rate $\gamma$ (note, $\Delta \equiv \frac{\partial^2}{\partial x_{1}^2} +  \frac{\partial^2}{\partial x_{2}^2}$); advection due to wind is defined by the advective derivative $\bm{v}_i(t) \cdot \nabla \equiv v_{x_1} \frac{\partial}{\partial x_1} + v_{x_2}\frac{\partial}{\partial x_2}$, where $\bm{v}_i(t) = (v_{x_1}, v_{x_2})'$ is the velocity vector at time $t$; deposition occurs at rate $\delta$; and $S_i(t)$ indicates the addition of \ce{SO4^2-} from external sources, in particular a chemical reaction described below.  Note that the advection-diffusion-deposition operator is acting on the concentration $y_i(t)$, while $S_i(t)$ occurs independent of $y_i(t)$. Finally, $\sigma \, \xi(\bm{s}, t)$ is a space-time white noise process with variance $\sigma^2$, capturing space-time varying sources and sinks of \ce{SO4^2-} that are otherwise unaccounted for in the model.

The predominant source of atmospheric \ce{SO4^2-} is thought to be sulfur dioxide (\ce{SO2}) emissions from coal-fired power plants \citep{Massetti2017}.  Therefore, we assume that \ce{SO4^2-}  is introduced to the system through a reaction term which depends on the local concentration of \ce{SO2}.  We also make the assumptions that the wind velocity vector $\bm{v}_i(t)$ is well-approximated by the annual average wind velocity (i.e., $\bm{v}_i$ is constant in time), and for computational convenience, that the advection-diffusion process for \ce{SO2} is the same as for \ce{SO4^2-}.  Thus, letting $z_i(t)$ denote the \ce{SO2} concentration, the process is defined by the coupled system of equations, 
\begin{align}
d y_i(t) & = \bigg[ \big( \gamma \Delta + \alpha \bm{v}_i \cdot \nabla - \delta \big) y_i(t) + \eta \, z_i(t) \bigg] dt +  \sigma \, \xi(\bm{s}, t), \label{eq:coupledSO4} \\
d z_i(t) & = \bigg[ \big( \gamma \Delta + \alpha \bm{v}_i \cdot \nabla - \eta \big) z_i(t) + \beta X_i \bigg] dt. \label{eq:coupledSO2}
\end{align}
Note that $\eta$ in \eqref{eq:coupledSO2} can be interpreted as the rate at which \ce{SO2} reacts into \ce{SO4^2-}. 
Lastly, $\beta X_i$ denotes the rate of emission of \ce{SO2} into the system from a coal-fired power plant located at $\bm{s}_i$, with annual \ce{SO2} emissions (in tons) represented by $X_i$. 

Together, Equations~\ref{eq:coupledSO4} and  \ref{eq:coupledSO2} define the space-time process from which we build a mechanistic spatial model for $\bm{Y_t} = (Y_{1}(t), \dots, Y_{n}(t))'$, the vector of observations of atmospheric \ce{SO4} observed at time $t$.  This model is sufficiently flexible to accommodate $\bm{Y_t}$ when viewed as either a temporal snapshot of the transient or stationary distribution of the generative process.  Furthermore, our model can be extended to spatial observations of the process averaged over a time interval $[0, T]$, defined as $\bm{V_T} = (V_{1}(T), \dots, V_{n}(T))'$, where $V_i(T) = \frac{1}{T} \int_{0}^{T} Y_i(s) ds$.   The latter includes observations of annual average sulfate concentrations, such as the \ce{SO4} observations displayed in Figure~\ref{subfig:emissions}.   For all three types of spatial data, the underlying atmospheric space-time dynamics play a critical role in the construction of the spatial model's mean and covariance structure.


\section{Constructing Spatial Models from Spatio-Temporal Ornstein-Uhlenbeck Processes}
\label{sec:spacetoOU}

We construct spatial models for a general class of spatio-temporal processes, such as the sulfate process presented in Section~\ref{sec:ctm}.  To do so, we approximate the process dynamics with a multivariate Ornstein-Uhlenbeck (OU) process, and leverage its distributional properties to specify novel probability models for data that are either a snapshot or an average over time of the OU process.  For clarity, we consider spatial models for spatio-temporal processes defined on continuous space; mechanistic spatial models for processes defined on discrete space can be constructed in a similar manner.

\subsection{Spatio-Temporal Process as an OU Process}
\label{subsec:STtoOU}

In continuous space and time ($\bm{s} \in \mathcal{D}$, $t \geq 0$), let $y_{\bm{s}}(t)$ denote the process of interest, and consider the class of stochastic processes defined by
\begin{equation}
d y_{\bm{s}}(t) = \bigg(-\mathcal{A}_{\bm{s}}(\bm{\theta}) y_{\bm{s}}(t) + m_{\bm{s}}(\bm{\theta})\bigg) dt + \mathcal{B}_{\bm{s}}(\bm{\theta}) \xi(\bm{s}, t), \label{eq:contprocess}
\end{equation}
where $\mathcal{A}_{\bm{s}}(\bm{\theta})$ denotes a linear differential operator (e.g., advection-diffusion), $m_{\bm{s}}(\bm{\theta})$ is a reaction term, and $\mathcal{B}_{\bm{s}}(\bm{\theta}) \xi(\bm{s}, t)$ is a space-time white noise process, with process variance defined by the real-valued function $\mathcal{B}_{\bm{s}}(\bm{\theta})$.  Each component in \eqref{eq:contprocess} may be parameterized by a vector of process parameters, $\bm{\theta}$. For notational convenience, the explicit parameterization denoted by $\bm{\theta}$ is omitted in the remainder of this section. For now, we assume that $\mathcal{A}_{\bm{s}}$, $m_{\bm{s}}$, and $\mathcal{B}_{\bm{s}}$ may vary in space, however, they are constant in time.  Equation~\eqref{eq:contprocess} encompasses a wide class of linear dynamical systems, including the mechanistic representation of sulfate transport from Section~\ref{sec:ctm}.

Although the continuous-space formulation of \eqref{eq:contprocess} is appropriate from a scientific perspective, its distributional properties are challenging to understand. This is especially true when $\xi(\bm{s}, t)$ varies across space and time. Consequently, it is convenient to consider a \textit{discrete space approximation} to \eqref{eq:contprocess}. Such an approximation can be constructed by first discretizing the continuous surface of interest, $\mathcal{D}$, into a finite collection of points, $\mathcal{S} = \{\bm{s}_1, \dots, \bm{s}_n\}$, and then approximating the linear operator $\mathcal{A}$ with a matrix operator $\bm{A}$, which is defined with respect to $\mathcal{S}$. A variety of numerical schemes have been developed to facilitate these approximations, including finite difference (FDM), finite volume (FVM), and finite element methods (FEM), and the relative merits and implementation details of each method are often problem-specific \citep{LangtangenLinge2017, Versteeg2007, Johnson2009}. As a result, we focus our discussion on the distributional consequences of discretizing the space-time process defined in \eqref{eq:contprocess}; additional discretization details, including an example discretization of an advection-diffusion-deposition process, can be found in the Supplementary Materials and the references therein.

By moving to discrete space, we can approximate the stochastic process in \eqref{eq:contprocess} with a multivariate It\^o stochastic differential equation (SDE).  Importantly, the stochastic process is now defined with respect to the discretization, $\mathcal{S}$, and $y_{\bm{s}}(t)$, $\mathcal{A}_{\bm{s}}$, $m_{\bm{s}}$, and $\mathcal{B}_{\bm{s}} \; \xi(\bm{s}, t)$ are replaced with their discrete counterparts. The continuous spatial surface, $y_{\bm{s}}(t)$, is restricted to an $n \times 1$ response vector $\bm{y}_t = (y_{1t}, \dots, y_{nt})'$, where $y_{it} \equiv y_{\bm{s}_i}(t)$, $\bm{s}_i \in \mathcal{S}$. Similarly, the surface of sources/sinks, $m_{\bm{s}}$, is replaced with a vector, $\bm{m} = (m_{\bm{s}_1}, \dots, m_{\bm{s}_n})'$, $\mathcal{B}_{\bm{s}}$ is replaced with an $n \times n$ matrix, $\bm{B}$, and the linear operator $\mathcal{A}_{\bm{s}}$ is replaced with its aforementioned sparse matrix approximation, $\bm{A}$. Finally, the space-time varying white noise process, $\xi(\bm{s}, t)$, is restricted to the discretized space, $\mathcal{S}$. Thus, we replace $\xi(\bm{s}, t)$ with an $n \times 1$ vector of Gaussian white noise, $d \bm{W}_t$ (i.e., $d \bm{W}_t$ is the distributional derivative of $\bm{W}_t$, where $\bm{W}_t = (W_1(t), \dots, W_n(t))'$ is $n$-dimensional Brownian motion, with $W_i(t)$ corresponding to location $\bm{s}_i \in \mathcal{S}$). Substituting these discrete approximations into Equation~\eqref{eq:contprocess} gives us 
\begin{equation}
d \bm{y}_t = \big( -\bm{A} \bm{y}_t + \bm{m} \big) dt + \sigma \bm{B} d \bm{W}_t \label{eq:OU}.
\end{equation}
We've approximated the continuous-space process \eqref{eq:contprocess} with a stochastic differential equation (SDE) \eqref{eq:OU} defined on a set of discrete spatial locations, $\mathcal{S}$; the process defined by this SDE is a multivariate Ornstein-Uhlenbeck process.

\subsection{Distributional Properties of the OU Process}

With origins firmly rooted in statistical mechanics \citep{OU1930, Jacobsen1996}, the Ornstein-Uhlenbeck (OU) process has been used in a diverse array of fields, including financial mathematics \citep{Eberlein1999}, infectious disease epidemiology \citep{Allen2016}, and evolutionary biology \citep{Bartoszek2017}.  The utility of the OU process can be attributed to its unique distributional properties.  Apart from white noise, it is the only continuous stochastic process that is simultaneously Gaussian, Markov, and stationary \citep{Doob1942}.  Consequently, the process is completely defined through its initial value and Gaussian transition probability distributions.  In addition, the process-dependent drift term in \eqref{eq:OU} results in a characteristic mean-reverting property.

Despite its prevalence in the physical sciences, the OU process has, to our knowledge, been relatively under-represented in the spatial statistics literature.  Notable exceptions include its use as the infinitesimal generator of the spatio-temporal Gaussian blurring model derived by \citet{Brown2000}, and as the stochastic intensity of the log-Gaussian Cox process presented in \citet{BrixDiggle2001}.  In both cases, the modeling efforts were focused on spatio-temporal data; the modeling of strictly spatial data (i.e., data with no temporal replication) via an OU process has not previously been addressed.  

\subsection{Spatial Models for Snapshot Data}

Assume that the initial state, $\bm{y}_0$, of the OU process $(\bm{y}_t)_{t \geq 0}$ in \eqref{eq:OU} is known.  Then, for a nonsingular matrix $\bm{A}$, $(\bm{y}_t)_{t \geq 0}$ is a Gaussian process with solution at time $t$ given by
\begin{equation}
\bm{y}_t = e^{-\bm{A} t} \bm{y}_0 + (\bm{I} - e^{-\bm{A} t}) \bm{A}^{-1} \bm{m} + \int_{0}^{t} \sigma e^{-\bm{A} (t - s)} \bm{B} d \bm{W}_s, \label{eq:OUsolution}
\end{equation}
where $e^{-\bm{A}t}$ denotes a matrix exponential and $\int_{0}^{t} \sigma e^{-\bm{A} (t - s)} \bm{B} d \bm{W}_s$ is a multivariate It\^o integral \citep{Oksendal2003}.  Consequently, the \textit{transient distribution} of the process at time $t$ is 
\begin{equation}
\bm{y}_t | \bm{y}_0 \sim N\bigg( e^{-\bm{A} t} \bm{y}_0 + (\bm{I} - e^{-\bm{A} t}) \bm{A}^{-1} \bm{m}, \int_{0}^{t} \sigma^2 e^{-\bm{A} (t - s)} \bm{B} \bm{B}' e^{-\bm{A}' (t - s)} ds \bigg). \label{eq:OUtransient}
\end{equation}
This result follows directly from \eqref{eq:OUsolution} \citep{Gardiner2004}.

The transient distribution presented in \eqref{eq:OUtransient} defines a spatial model for the spatio-temporal process assumed in \eqref{eq:contprocess}.    In particular, it assumes that the process was observed at some known time $t$, and that information about the initial state $\bm{y}_0$ is either known or desired.  For example, during the outbreak of an invasive species \citep{Hooten2007}, the initial state represents the introduction of the agent into the system, and there may be known information about this introduction.  By incorporating this information into \eqref{eq:OUtransient}, inference on the process dynamics -- as defined by $\bm{A}$, $\bm{m}$, and $\sigma \bm{B}$ -- may be obtained from spatial data alone.  In some cases inference about $\bm{y}_0$, such as the location or time the introduction most likely occurred, may itself be inferred from spatial data \citep{Hefley2017b}.

However, in many systems little is known about the past, and directly characterizing $\bm{y}_0$ is difficult or impossible.  For such scenarios, we instead consider the \textit{stationary (time-limiting) distribution} of the OU process, which we denote as 
\begin{equation}
\bm{y}_{\infty} = \lim_{t \to \infty} f(\bm{y}_t | \bm{y}_0), \label{eq:statdef}
\end{equation}
where $f(\bm{y}_t | \bm{y}_0)$ is the transient distribution in \eqref{eq:OUtransient}.  Assuming $\bm{A}$ has only eigenvalues with positive real part, the stationary distribution of \eqref{eq:OU} exists \citep{Gardiner2004} and is given as
\begin{equation}
\bm{y}_{\infty} \sim N \bigg( \bm{A}^{-1} \bm{m}, \bm{\Sigma} \bigg). \label{eq:OUstat}
\end{equation}
Here $\bm{\Sigma}$ is the solution to the continuous Lyapunov equation,
\begin{equation}
\bm{A} \bm{\Sigma} + \bm{\Sigma} \bm{A'} = \sigma^2 \bm{B} \bm{B}'. \label{eq:Lyap}
\end{equation}

Solving the Lyapunov equation can quickly become computationally expensive as the size of $\bm{A}_{n \times n}$ increases.  For example, the common Bartels-Stewart algorithm \citep{Bartels1972} requires $\mathcal{O}(n^3)$ floating point operations, although for low-rank $\bm{B}$ faster methods are available \citep{Simoncini2007}.  In the case that $\bm{A} = \bm{A'}$ (i.e., $\bm{A}$ is symmetric), the solution to \eqref{eq:Lyap} is simply $\bm{\Sigma} =  \sigma^2 \bm{A}^{-1} \bm{B} \bm{B}' / 2$.  Furthermore, if we assume an uncorrelated background driving process in \eqref{eq:OU}, corresponding to the common assumption $\bm{B} = \bm{I}$, the stationary distribution of \eqref{eq:OU} is simply
\begin{equation}
\bm{y}_{\infty} \sim N \bigg(\bm{A}^{-1} \bm{m}, \frac{\sigma^2}{2} \bm{A}^{-1} \bigg). \label{eq:OUstatsimple}
\end{equation}
Thus, when the linear operator $\bm{A}$ is symmetric, the stationary distribution reduces to a spatial CAR model with precision $\frac{2}{\sigma^2} \bm{A}$.

Together, the transient and stationary distributions given in \eqref{eq:OUtransient} and \eqref{eq:OUstat} form a class of mechanistic models for spatial data observed as a \textit{snapshot} of a larger space-time process.  In this ``snapshot'' observation, the time-scale of the space-time process should be much larger than the observational window in which the data were collected.  For example, both daily sea surface temperatures and landscape genetics data might be considered realistic snapshots of their respective generating processes.  In cases where additional information is known or desired about the initial state of the system, the data should be modeled using the transient distribution \eqref{eq:OUtransient} of the OU process.  When the initial state is unknown, or the process can reasonably be assumed to have reached stationarity, the data should be modeled using the stationary distribution \eqref{eq:OUstat}.

\subsection{Spatial Models for Time-Averaged Data}  
\label{subsec:averaged}

Often data are collected and aggregated over time and space. For example, the average daily concentration of fine particulate matter in the atmosphere over the course of a year, the total monthly precipitation, or the number of reported cases of a disease over the course of a year. In each of these cases, and in many more scenarios in ecology and environmental science, time-aggregated spatial data arise from processes with well-understood dynamics.  The properties of the OU process allow us to extend our mechanistic spatial model to time-averaged spatial data.  We show that the length of time over which spatial observations are aggregated has important implications on the second-order structure of the model.   

Once again, consider the OU process defined by \eqref{eq:OU}, and assume that $\bm{y}_0$ is a draw from the stationary distribution.  Then $(\bm{y}_t)_{t \geq 0}$ is a Gaussian process with mean
\begin{equation}
\bm{\mu}(t) \equiv \bm{A}^{-1} \bm{m} \label{eq:GPmean}
\end{equation}
and covariance
\begin{equation}
k(s, t) \equiv Cov(\bm{y}_s, \bm{y}_t) = \begin{cases}
      \bm{\Sigma} e^{-\bm{A}' (t - s)}, & s < t\\
      e^{-\bm{A} (s - t)} \bm{\Sigma}  & t < s,
    \end{cases}      \label{eq:GPcov}
\end{equation}
where $\bm{\Sigma}$ is the solution to the Lyapunov equation \eqref{eq:Lyap}.  If we then integrate this Gaussian process over a time window $[0, T]$ of length $T$, the resulting \textit{time-averaged process} $(\bm{v}_T)_{T \geq 0}$ is also Gaussian, with
\begin{equation}
\bm{v}_T = \frac{1}{T} \int_{0}^{T} \bm{y}_s ds \sim N \big(\bm{A}^{-1} \bm{m}, \bm{\Psi} \big). \label{eq:avg}
\end{equation} 
The covariance matrix, $\bm{\Psi}$, takes the form
\begin{equation}
\bm{\Psi} = \frac{\sigma^2}{T} \bigg(\bm{A}' (\bm{B}\bm{B}')^{-1} \bm{A} \bigg)^{-1} - \frac{1}{T^2} \bigg[ \bm{\Sigma} (\bm{I} - e^{-\bm{A}' T}) (\bm{A}')^{-2} + (\bm{I} - e^{-\bm{A}T}) \bm{A}^{-2} \bm{\Sigma} \bigg]. \label{eq:fullcovar}
\end{equation}
This result is an extension of Doob's (1942) derivation of the distribution of displacements for a univariate OU process.  Mathematical details are included in Appendix A.

The covariance $\bm{\Psi}$ of the time-averaged process can naturally be broken into two component parts, 
\begin{equation}
\bm{\Phi} =  \frac{\sigma^2}{T} \big(\bm{A}' (\bm{B}\bm{B}')^{-1} \bm{A} \big)^{-1}, \label{eq:covapprox}
\end{equation}
and the remainder, $\bm{E} = \bm{\Psi} - \bm{\Phi}$.  Notice that, from a computational perspective, $\bm{E}$ requires both the evaluation of a matrix exponential as well as the solution to the Lyapunov equation, two expensive tasks each of order $\mathcal{O}(n^3)$ flops.  In contrast, the operator matrix $\bm{A}$ is often sparse -- a common property of most numerical approximation methods.  If we again assume $\bm{B} = \bm{I}$, then $\bm{\Phi}$ now has the computationally convenient form
\begin{equation}
\bm{\Phi} = \frac{\sigma^2}{T} \big(\bm{A}' \bm{A} \big)^{-1}. \label{eq:betterapprox}
\end{equation}
Thus, it is prudent to ask under what scenarios $\bm{\Phi}$ accurately approximates $\bm{\Psi}$ in \eqref{eq:fullcovar}.

For example, consider a homogeneous diffusion process with constant rate of decay, $\delta$ -- a relatively simple model for the dispersal of air pollution about a source, or of species movement across a landscape.  This process can be approximated by the linear operator $\bm{A} = (\gamma \bm{D} + \delta \bm{I})$, where $\bm{D}$ is a second-order central difference matrix and $\gamma$ a constant rate of diffusion.  For spatial data averaged over a given interval of time, $[0, T]$, how well does $\bm{\Phi}$ approximate the covariance matrix $\bm{\Psi}$ in \eqref{eq:avg}? In this case, we show (Appendix B) that 
\begin{equation}
\vert \vert \bm{E} \vert \vert_{2} = \vert \vert \bm{\Psi} - \bm{\Phi} \vert \vert_{2} \leq \frac{1}{T^2 \delta^3} (1 - e^{-\delta T}).
\end{equation}
Thus, for large $\delta T$, the remainder $\bm{E}$ goes to zero in the $L_2$ norm, and we have $\bm{\Psi} \approx \bm{\Phi}$.  More generally, $\bm{\Psi}$ is well-approximated by $\bm{\Phi}$ for any aggregated space-time process in which the observation length $T$ is sufficiently large to allow for repeated turnover of the system dynamics.  In such scenarios, the covariance approximation in \eqref{eq:betterapprox} is in the form of a SAR model with precision matrix proportional to $\bm{A}'\bm{A}$. 

\subsection{Summary of OU Spatial Models}

In summary, we have constructed a class of mechanistic spatial models applicable to spatial data observed from a variety of dynamical systems.  These models are especially useful for data observed from systems that are well-approximated by (locally) linear systems of equations, such as the reaction--convection--diffusion equation (potential applications include invasive species growth, population migration, pollution transport, cellular morphogenesis, etc.) or the linear shallow water equations (e.g., tropical meteorological systems, tsunami wave propagation).  After approximating these systems with a multivariate OU process, we obtain a familiar Gaussian process framework for modeling spatial data.  This approach has two main advantages over existing phenomenological spatial models.  First, the underlying physical processes determine the spatial model's mean and covariance structure, implying that inference may be obtained on important components of the space-time dynamics from spatial data alone.  This is especially useful for such systems where it is infeasible to observe spatial observations repeatedly in time.  Second, the model is flexible enough to be used with three types of spatial data: observations made over a short temporal time frame (i.e., `snapshot' data) from either the process's transient or stationary distribution, and time-averaged spatial data assumed to be observed after the process has reached stationarity.   As demonstrated, the temporal support of the spatial data has important consequences on the covariance structure of the model.  In special cases, snapshot observations can be modeled with a Gaussian process with a CAR covariance structure with precision matrix proportional to the (sparse) matrix operator $\bm{A}$, while the covariance structure for time-averaged observations reduces to a SAR specification with precision proportional to $\bm{A'A}$.

\section{Analysis of Atmospheric Sulfate}
\label{sec:analysis}

We now apply this novel class of mechanistic spatial models to the 2011 annual sulfate data introduced in Section~\ref{sec:data}.  To begin, we approximate the coupled, mechanistic representation of \ce{SO4^2-}-\ce{SO2} in (\ref{eq:coupledSO4}, \ref{eq:coupledSO2}) with an OU process \eqref{eq:OU} discretized on a $70 \times 116$ regular grid (cell area $\sim$ \SI{200}{\km\squared}).  Letting $\bm{y}_t$ and $\bm{z}_t$ denote the vectorized \ce{SO4^{2-}} and \ce{SO2} concentrations resulting from coal-fired power plant emissions, the discrete-space process is defined by
\begin{align}
\bm{y}_t & = \bigg[ - (\gamma \bm{D} + \alpha \bm{C} + \delta \bm{I}) \bm{y}_t + \eta \bm{z}_t \bigg] dt + \sigma d \bm{W}_t, \label{eq:sdeSO4}\\
\bm{z}_t & = \bigg[ - (\gamma \bm{D} + \alpha \bm{C} + \eta \bm{I}) \bm{z}_t + \beta \bm{X} \bigg] dt. \label{eq:sdeSO2}
\end{align}
Here, $-(\gamma \bm{D} + \alpha \bm{C})$ is the finite volume method approximation of an advection-diffusion process caused by wind, where $\bm{D}$ is a second-order central difference matrix and $\bm{C}$ is a first-order upwind discretization, with edge flux assigned the interpolated 2011 average wind velocity (10 \si{m}).  See the Supplementary Material for more discretization details.  

From a meteorological perspective, $\gamma \bm{D}$ represents transport due to sub-annual wind variability, while $\alpha \bm{C}$ denotes advection due to annual wind velocity.  As before, $\delta$ is the rate of \ce{SO4^2-} deposition, while $\eta$ is the rate at which \ce{SO2} attributed to coal-fired power plants ($\bm{z_t}$) reacts into \ce{SO4^2-}. Thus, \eqref{eq:sdeSO4} corresponds to the SDE in \eqref{eq:OU}, with $\bm{A} = (\gamma \bm{D} + \alpha \bm{C} + \delta \bm{I})$, $\bm{m}(t) = \eta \bm{z}_t$, and $\bm{B} = \bm{I}$. The coupled (deterministic) differential equation \eqref{eq:sdeSO2} explicitly links coal-fired power plant emissions of \ce{SO2} ($\beta \bm{X}$) to the sulfate concentration, $\bm{y}_t$, through a similarly defined advection-diffusion process. Notably, the two processes (\ref{eq:sdeSO4}, \ref{eq:sdeSO2}) are linked through the reaction of \ce{SO2} emissions into \ce{SO4^2-} at rate $\eta$.

The 2011 observations used in our analysis are of \textit{annual} average sulfate concentrations.  Given the speed at which pollution transport occurs \citep{SeinfeldPandis2016}, it is ill-advised to consider such data as a temporal snapshot.  Instead, the time-averaged spatial model presented in Section~\ref{subsec:averaged} is an appropriate model for these data.  Let $\bm{V}_{T} = \frac{1}{T} \int_{0}^{T} \bm{y}_t dt$ denote the averaged process over time $T$ (in years). Similarly, let 
\begin{equation}
\bm{Z}_{\bm{\theta}, \bm{X}} \equiv \lim_{t \to \infty} \bm{z}_{t} = (\gamma \bm{D} + \alpha \bm{C} + \eta \bm{I})^{-1} (\beta \bm{X}) \label{eq:steadySO2}
\end{equation} 
denote the steady-state solution of the deterministic \ce{SO2} process defined in \eqref{eq:sdeSO2}, as determined by process parameters, $\bm{\theta}$, and the known \ce{SO2} power plant emission levels, $\bm{X}$. Then, $\bm{V}_T$ may be appropriately modeled via the time-averaged mechanistic model \eqref{eq:avg},
\begin{equation}
\bm{V}_T \sim N\big( \bm{\mu}(\bm{\theta, X}), \; \bm{\Sigma}_{\bm{\theta}} \big),  \label{eq:avgmodel}
\end{equation}
where 
\begin{equation}
\bm{\mu}(\bm{\theta, X}) = \bm{A}_{\bm{\theta}}^{-1} ( \eta \; \bm{Z}_{\bm{\theta}, \bm{X}}) \label{eq:avgmean}
\end{equation}
 and 
 \begin{equation}
 \bm{\Sigma}_{\bm{\theta}} = \frac{\sigma^2}{T} (\bm{A_{\theta}}' \bm{A_{\theta}})^{-1}. \label{eq:avgcovar}
 \end{equation}

This model is over-parameterized, as both $\gamma$ and $\delta$ (components of $\bm{A_{\theta}}$) control the spatial smoothness of the process.  However, the rate of atmospheric deposition of \ce{SO4^2-} is quite fast (i.e., less than a week \citep{SeinfeldPandis2016}), and so we fix $\delta$ at 50.  With a time scale of $T = 1$ year, this implies a rate of turnover consistent with the known process \citep{SeinfeldPandis2016}. Thus, our mechanistic model for a complex space-time process -- annual average \ce{SO4^2-} concentrations attributed to coal-fired power plant emissions of \ce{SO2} -- has been reduced to a conveniently simple model, a familiar Gaussian process with mean $\bm{\mu}(\bm{\theta, X})$ and SAR precision matrix  $\bm{\Sigma}_{\bm{\theta}}^{-1}$ defined via the sparse matrix operator $\bm{A_{\theta}}$ and the vector of \ce{SO2} power plant emissions, $\bm{X}$.

\subsection{Inference}

We constructed a Markov chain Monte Carlo (MCMC) algorithm to sample from the posterior distribution of model parameters, given the observed 2011 \ce{SO4^2-} concentrations and coal-fired power plant emissions data.  Updates for all model parameters except $\beta$ were obtained using a random walk Metropolis step, with likelihood as defined in \eqref{eq:avgmodel} and priors given in Table~\ref{tab:params}.  For $\beta$, a full conditional Gibbs update was used.  Five chains were run with randomly selected starting values, and each chain was run for 150,000 iterations, with the first 25,000 samples discarded as burn-in.  Convergence was assessed qualitatively across the five chains.

	\begin{table}[ht]
	\caption {Parameter estimates for atmospheric \ce{SO4^2-} analysis. Unit of time $T = 1$ year.} \label{tab:params}
	\setlength{\tabcolsep}{12pt}
	\small 
	\centering
	\begin{tabular}{llrcl}
	  \toprule
	  Parameter & Interpretation & $\bm{\hat{\theta}} \approx$ E($\bm{\theta} | \bm{Y}$) & 95\% CI  & Prior\\ 
	  \midrule
	  $\gamma$  & rate of sub-annual wind transport &1510 & (1305, 1725) &  Half-Normal \\ 
	  $\alpha$ & rate of annual wind transport & 0.53 & (0.02, 1.87) & Half-Normal \\ 
	  $\eta$ & \ce{SO2} $\rightarrow$ \ce{SO4^2-} & 0.50 & (0.36, 0.69) & Exponential \\ 
	  $\beta$ & proportional rate of \ce{SO2} emission & 3.45 & (3.30, 3.71) & Half-Normal \\ 
	  $\sigma^2$ & B.M. process variance & 24,000 & (18100, 31000) & Exponential \\
	  $\delta$ & deposition of \ce{SO4^2-} & NA & NA & fixed at 50  \\ 
	   \bottomrule
	\end{tabular}
	\end{table}

Posterior mean estimates and 95\% equal-tailed credible intervals are included in Table~\ref{tab:params}.  Because \eqref{eq:coupledSO4} is a simple approximation to a complex process, not all parameter estimates have a clear scientific interpretation (e.g., $\beta$).  However, some general trends can be inferred.  The estimated rate of sub-annual wind transport, $\hat{\gamma} = 1510$, is much larger than the rate of advection due to annual average wind velocity, $\hat{\alpha} = 0.53$, implying that sub-annual variability in the wind field dominates observed \ce{SO4^2-} transport averaged over a year.  Furthermore, the estimate $\hat{\eta} = 0.53$ implies that the \ce{SO2} process acts as a smoothed source of \ce{SO4^2-}, centered about power plant locations.  Finally, given the estimates of $\hat{\gamma}$ and $\hat{\alpha}$, $\hat{\sigma}^2 = 24,000$ implies that the marginal variances of the averaged \ce{SO4^2-} process (i.e., the marginal variances of $\bm{\hat{\Sigma}}_{\hat{\bm{\theta}}}$ from \eqref{eq:avgmodel}) range from approximately $0.01$ to $0.1$.

\begin{figure}[ht]
\centering
    \begin{subfigure}[b]{0.4\textwidth}
        \includegraphics[width=\textwidth]{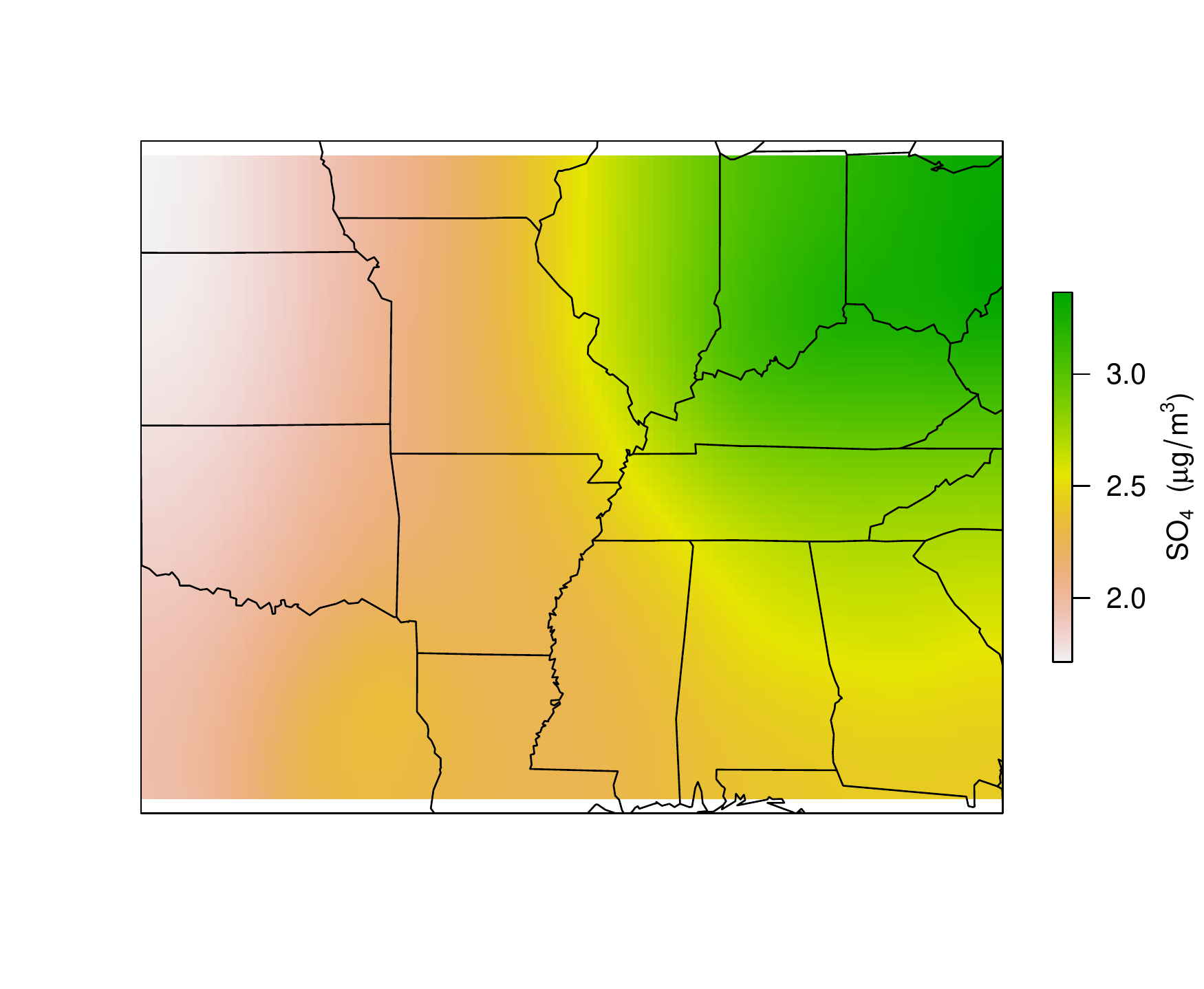}
        \caption{Estimated mean \ce{SO4^2-} from power plant emissions.}
        \label{subfig:postmean}
    \end{subfigure}
    ~ 
    \begin{subfigure}[b]{0.4\textwidth}
        \includegraphics[width=\textwidth]{2011emissions.pdf}
        \caption{Observed 2011 average \ce{SO4^2-} concentrations, with \ce{SO2} emissions sources.}
        \label{subfig:real2}
    \end{subfigure}
    \caption{A comparison of (a) the estimated mean \ce{SO4^2-} from power plant emissions with (b) the observed average 2011 sulfate concentrations.}
    \label{fig:2011so4}
\end{figure}

A qualitative comparison of the the observed 2011 data with the estimated mean annual \ce{SO4^2-} concentrations attributed to \ce{SO2} emissions (Figure~\ref{fig:2011so4}), calculated as 
\begin{equation}
\bm{\hat{V}} = \bm{A}^{-1}_{\hat{\bm{\theta}}} (\hat{\eta} \; \bm{Z}_{\hat{\bm{\theta}}, \bm{X}}), \;\;  \hat{\bm{\theta}} = E(\bm{\theta} | \bm{V}), \label{eq:meanest}
\end{equation}
provides additional insight regarding the inferred pollution dynamics. Recall that our dynamical system represents the annual average concentration of \ce{SO4^2-} ($\bm{V_T}$) due to \ce{SO2} emissions from coal-fired power plants ($\bm{X}$). All other sources of $\ce{SO4}$ are assumed to be captured in the random white-noise process in \eqref{eq:sdeSO4}. Thus, a comparison of the the observed 2011 average sulfate concentrations against the estimated mean surface of sulfate pollution due to power plant emissions, $\bm{\hat{V}}$, can be used to both highlight geographic areas where observed \ce{SO4^2-} is likely due to power plant emissions, while at the same time identify areas in which observed \ce{SO4^2-} can most likely be attributed to alternative emissions sources.

For example, we see that the estimated mean \ce{SO4^2-} attributed to \ce{SO2} power plant emissions (Figure~\ref{subfig:postmean}) is highest in the Ohio river valley, with a similar pocket of increased \ce{SO4^2-} in east Texas. These estimated regions of high \ce{SO4^2-} exposure correspond with the locations of the largest \ce{SO2} emitting power plants in the country (Figure~\ref{subfig:real2}). Our model also identifies areas where the observed 2011 \ce{SO4^2-} concentrations are most likely due to sources other than power plant \ce{SO2} emissions, particularly along the Mississippi River and the Gulf Coast. We hypothesize that the unexplained \ce{SO4^2-} in these areas is largely due to sulfur dioxide emissions from shipping traffic and heavy industrial activity \citep{Mostert2017}.  Additional differences between the estimated and observed \ce{SO4^2-} concentration can be attributed to random variation in \ce{SO4^2-} sources and sinks, which are included in the estimated covariance structure of our model.

\begin{figure}[ht]
\centering
    \begin{subfigure}[b]{0.3\textwidth}
        \includegraphics[width=\textwidth]{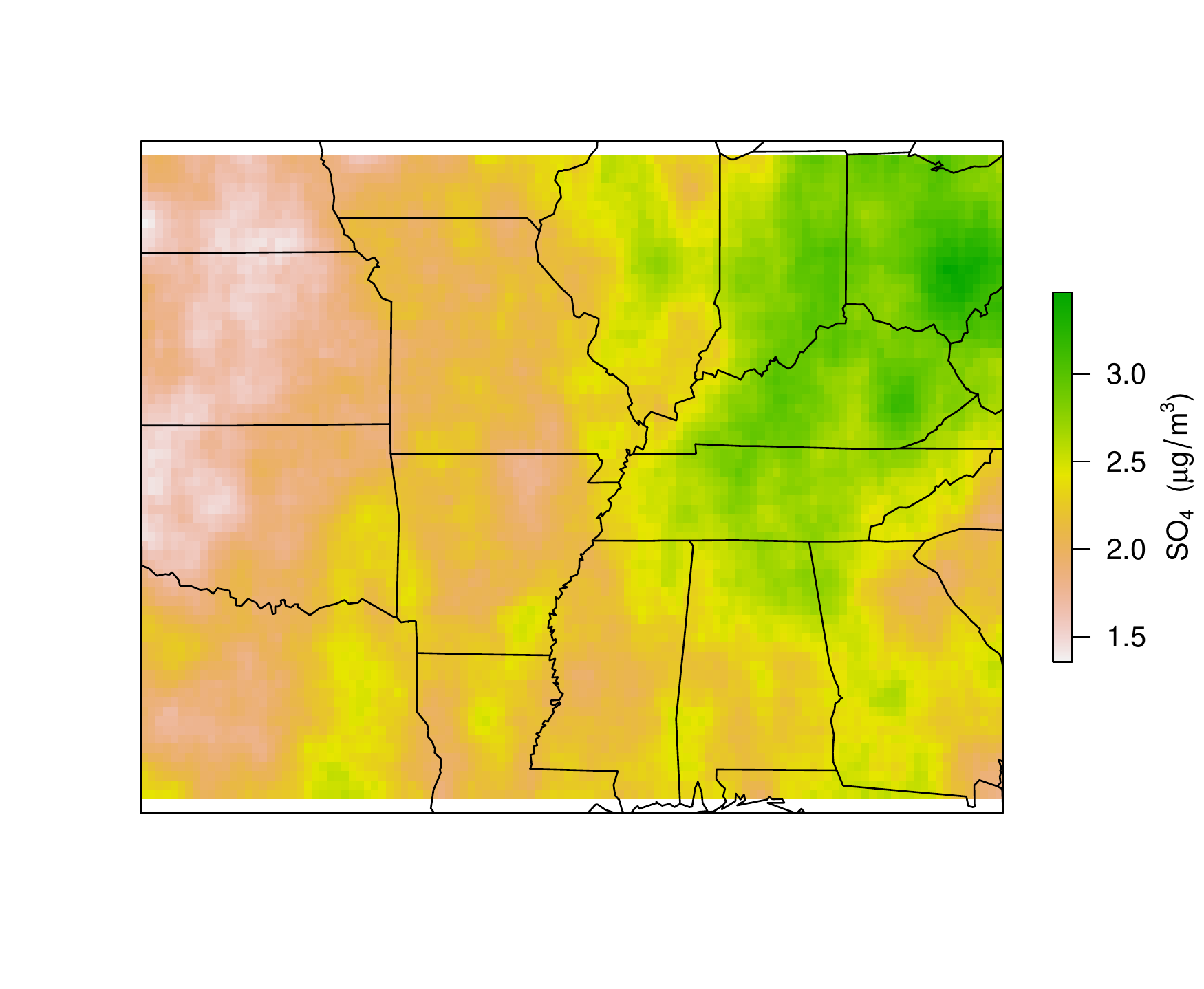}
        \caption{Averaged, coupled.}
        \label{subfig:simbest}
    \end{subfigure}
    ~ 
    \begin{subfigure}[b]{0.3\textwidth}
        \includegraphics[width=\textwidth]{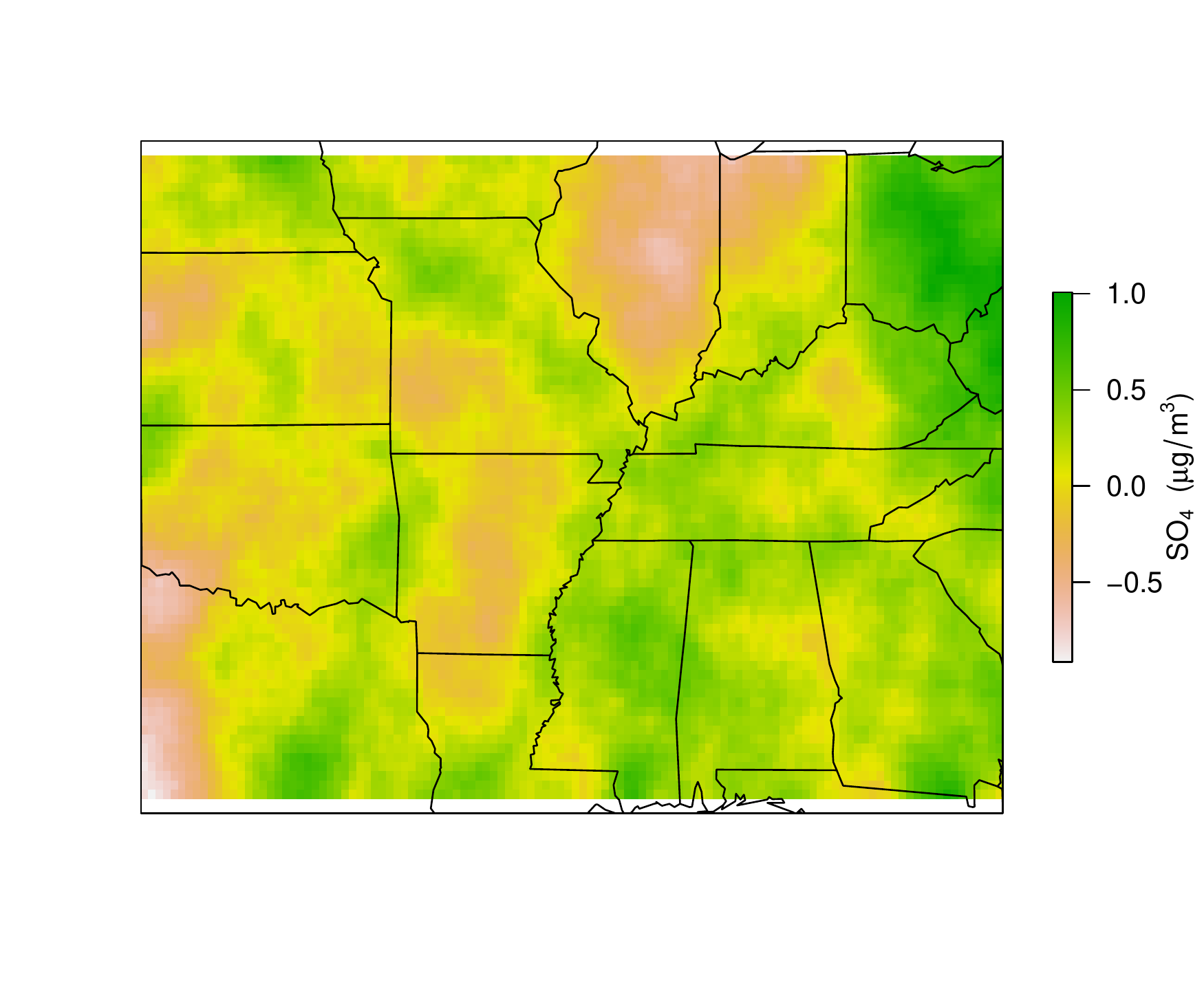}
        \caption{Averaged, uncoupled.}
        \label{subfig:simbad}
    \end{subfigure}
       ~
     \begin{subfigure}[b]{0.3\textwidth}
        \includegraphics[width=\textwidth]{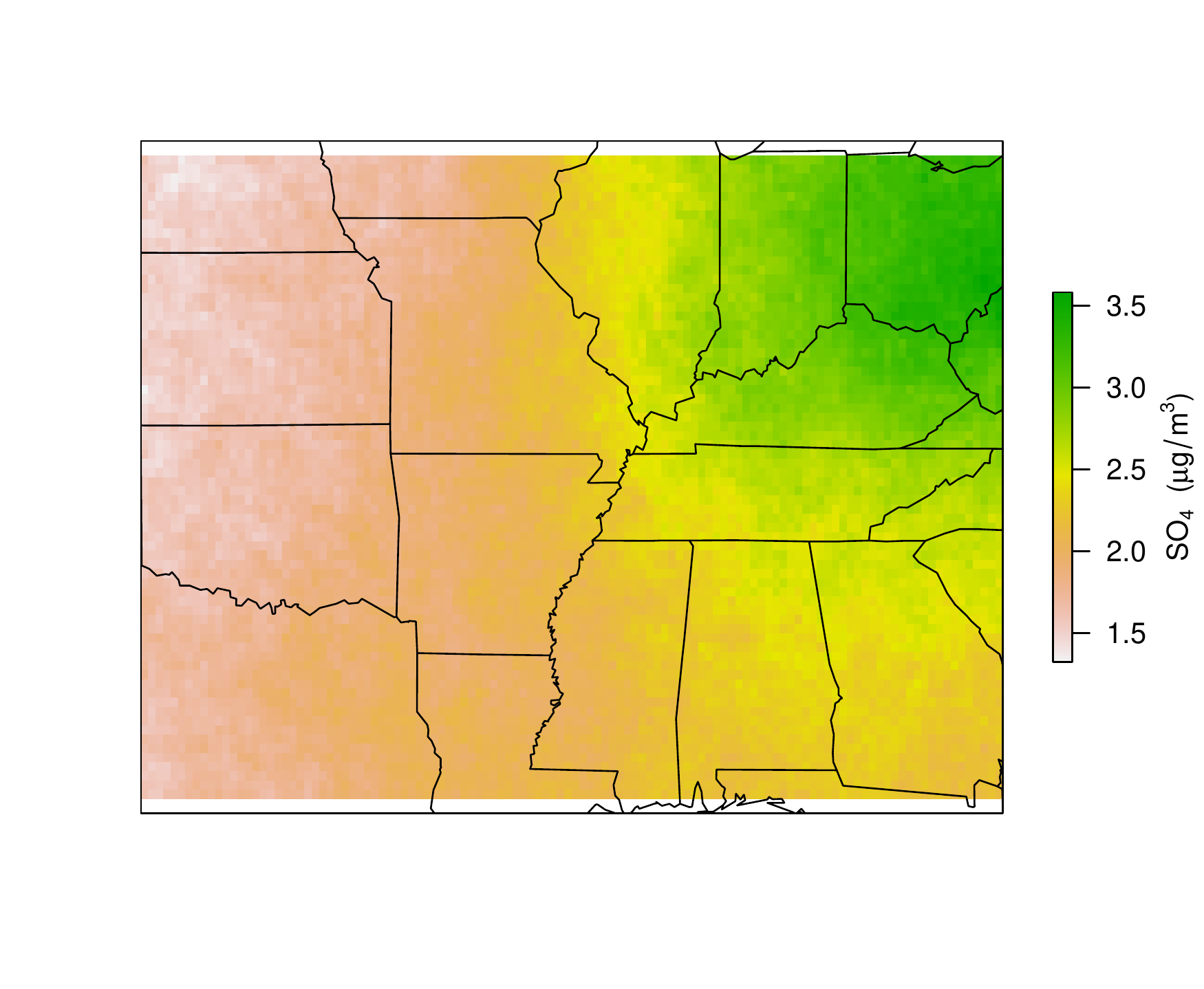}
        \caption{Snapshot, coupled.}
        \label{subfig:simsnap} 
    \end{subfigure}
    \caption{Simulated processes for the (a) time-averaged data with coupled SDE, (b) time-averaged data with uncoupled SDE, and (c) snapshot data with coupled SDE.  Compared with the observed sulfate concentrations, the time-averaged \ce{SO2} -- \ce{SO4^2-} model (a) performed best; the simplified time-average model (b) has an issue with scale, while the snapshot model (c) has poor second-order structure.}
    \label{fig:simexamples}
\end{figure}

For illustrative purposes, we compared the fitted model with two alternatives: a time-averaged spatial model \eqref{eq:avg} with power plant emissions now included as a direct source of \ce{SO4^2-}, and a stationary `snapshot' model \eqref{eq:OUstat} with the same generative process as Equation \eqref{eq:sdeSO4}.  Model fit was assessed based on the deviance information criterion (DIC, \cite{Spiegelhalter2002}), along with a qualitative assessment of posterior predictive draws from the process.  Both the simplified process model (i.e., no joint \ce{SO2} component) and the snapshot model (DIC = -47,370 and -43,140, respectively) were found to be inferior to the time-averaged model with coupled \ce{SO2} -- \ce{SO4^2-} (DIC = -47,790).  Similarly, a visual comparison of spatial fields sampled from the posterior distribution of each model (Figure~\ref{fig:simexamples}) reveals the superiority of the time-averaged, coupled system.  When compared with the observed 2011 sulfate concentrations (Figure~\ref{subfig:real2}), the snapshot model has a misspecified error structure (Figure~\ref{subfig:simsnap}), while the uncoupled process model has a problem with scale (Figure~\ref{subfig:simbad}).  In contrast, the time-averaged model with joint \ce{SO2} -- \ce{SO4^2-} matches the observed sulfate surface in both scale and spatial smoothness pattern (Figure~\ref{subfig:simbest}).  These results indicate that 1) the temporal support of the data (in this case, annual sulfate concentrations) should be carefully considered when choosing an appropriate spatial model, and that 2) careful effort should be spent verifying that the key dynamics of the underlying scientific process are included in the OU process construction. 

\subsection{Estimating Human Exposure to \ce{SO4^2-}}
\label{subsec:humanexposure}

An important advantage of mechanistic models for spatial data are their ability to provide probabilistic forecasts under alternative process scenarios.  This is especially useful when assessing the regulatory impacts of power plant emissions on observed air pollution, which is a task traditionally left to deterministic physical-chemical models with limited ability to characterize uncertainty.  For example, flue-gas desulfurization (FGD) technologies are used to remove (i.e., `scrub') \ce{SO2} from coal-fired power plant emissions.  Given the known impacts of \ce{SO4^2-} on human health (see Section~\ref{sec:data}), we consider which power plant facilities should be targeted with FGD systems in order to best reduce the overall human exposure to \ce{SO4^2-}.  

\begin{figure}[!ht]
\centering
\begin{subfigure}{.45\textwidth}
\centering
\includegraphics[width=\textwidth]{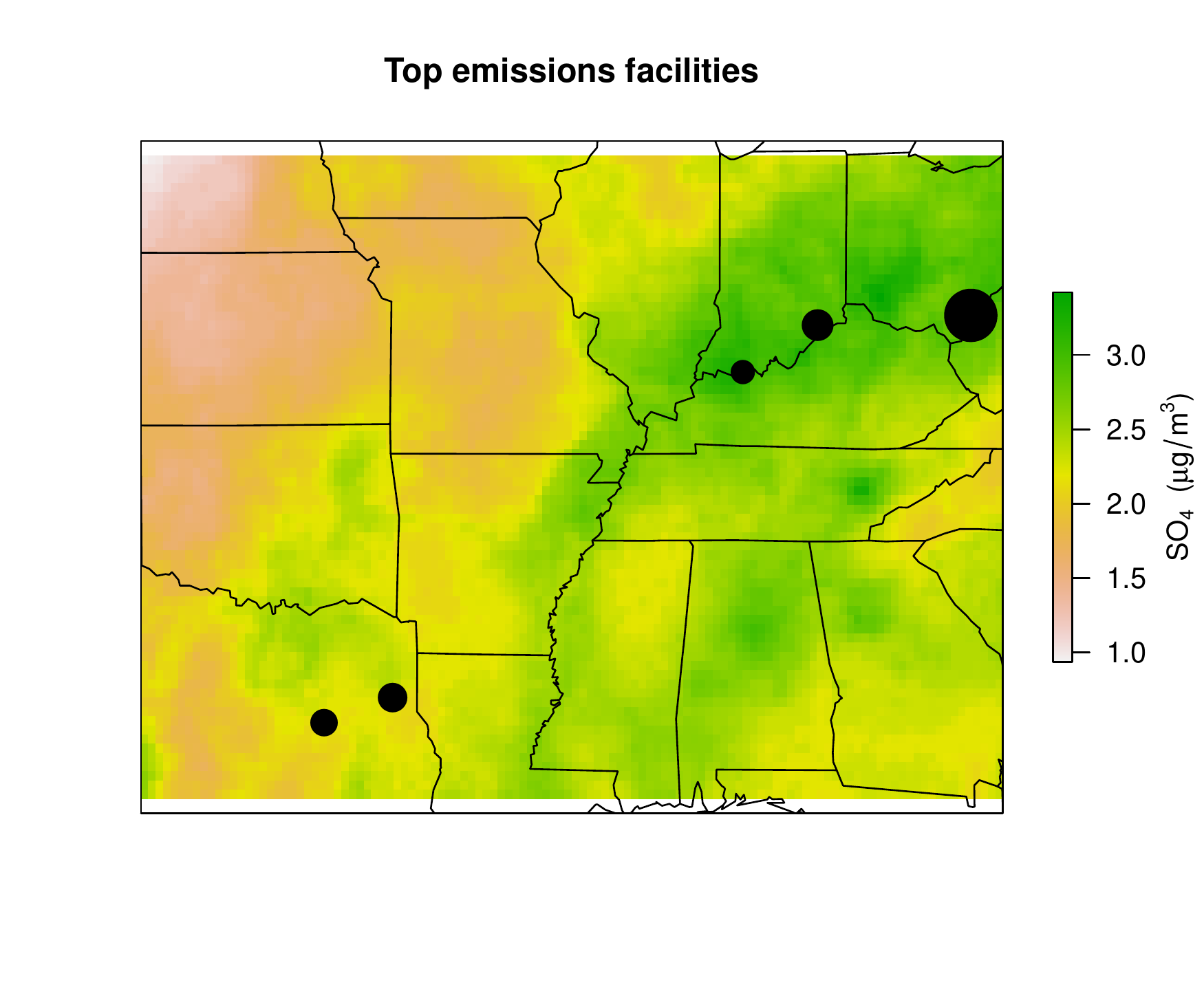}
\caption{Top five \ce{SO2} sources (tons/year).}
\label{subfig:top5}
\end{subfigure}%
\begin{subfigure}{.45\textwidth}
\centering
\includegraphics[width=\textwidth]{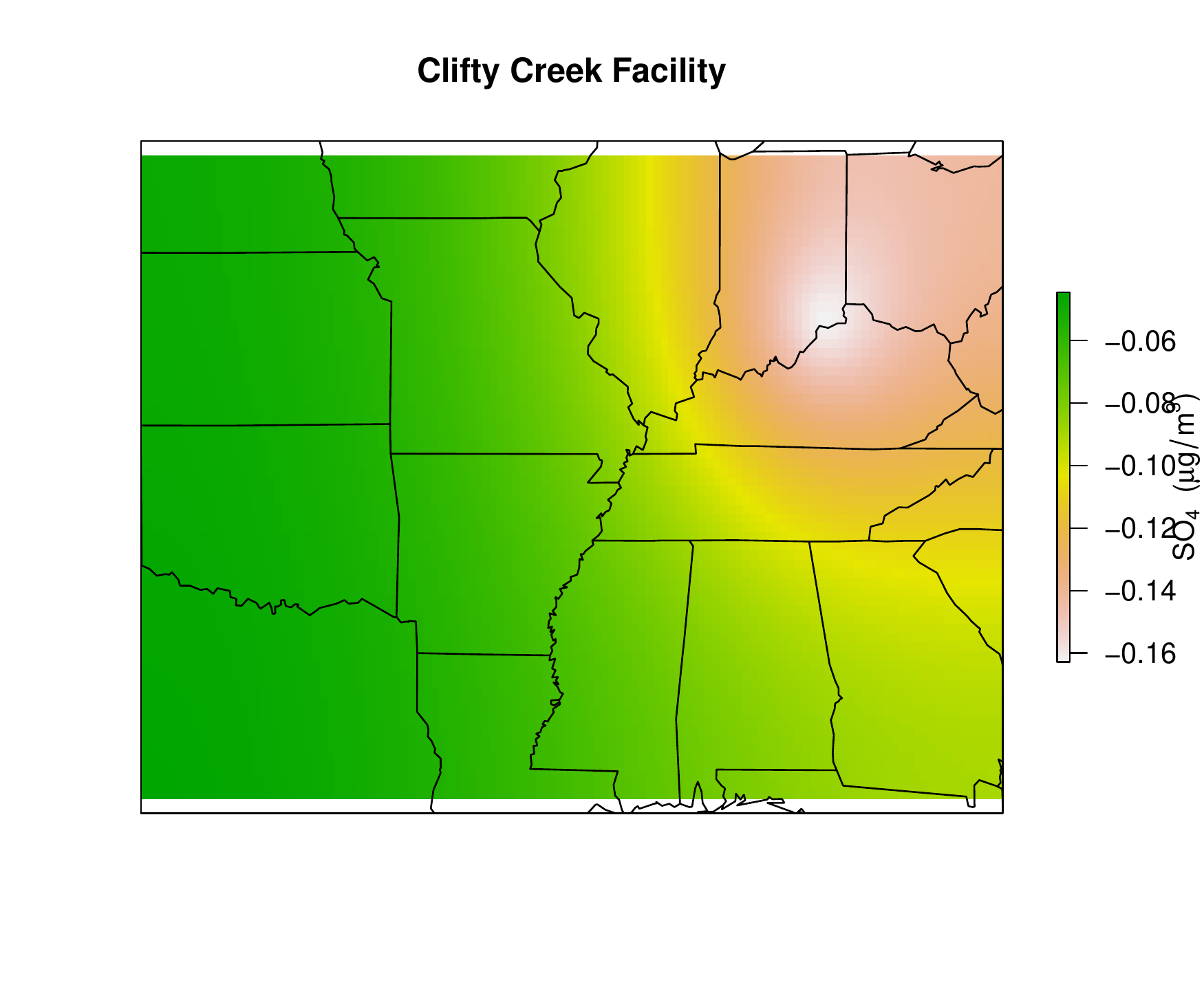}
\caption{Expected decrease in \ce{SO4^2-} after scrubber.}
\label{subfig:clifty}
\end{subfigure}\\[1ex]
\begin{subfigure}{.75\textwidth}
\centering
\includegraphics[width=\textwidth]{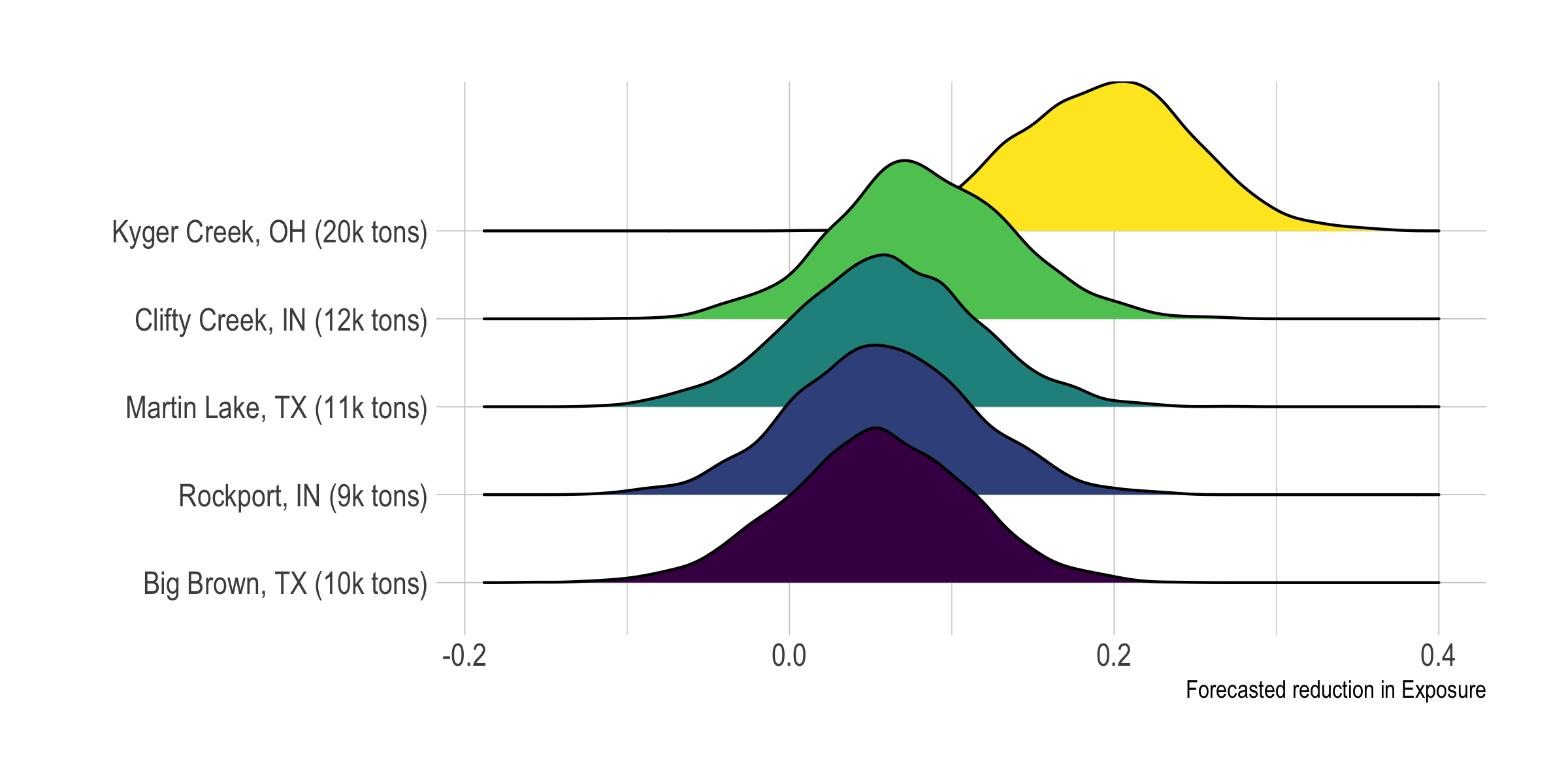}
\caption{Probabilistic forecasts of the average reduction in human exposure to \ce{SO4^2-} among five intervention scenarios; $(\cdot)$ denotes 2011 emissions totals (before FGD reduction).}
\label{subfig:exp-est}
\end{subfigure}
\caption{A regulatory assessment of FGD scrubber intervention on annual \ce{SO4^2-} exposure among the top five \ce{SO2} emitting facilities of 2011.}
\label{fig:exposure}
\end{figure}

As an example, consider the five largest power plant facilities from our 2011 analysis (Figure~\ref{subfig:top5}), based on the total amount of \ce{SO2} emitted per facility. The inferred process dynamics from the fitted model (see \eqref{eq:steadySO2} and \eqref{eq:avgmodel}) can be used to create probabilistic estimates of the expected decrease in 2011 \ce{SO4^2-} concentration if an FGD scrubber had been in place at each facility.  Let $\bm{X}^{*}_i$ denote the decrease in \ce{SO2} emitted from facility $i$ if an FGD scrubber had been in place in 2011. Using parameter samples from the posterior, $\bm{\theta}^{(k)} \sim \pi(\bm{\theta} | \bm{Y})$, we can simulate sample surfaces of the reduction in 2011 \ce{SO4^2-} after FGD scrubber implementation at facility $i$,
\begin{equation}
\hat{\bm{V}}^{(k)}_{i} | \bm{X}^{*}_i \sim N\big(\bm{\mu}(\bm{\theta}^{(k)}, \bm{X}^{*}_i), \bm{\Sigma}_{\theta^{(k)}}\big), \label{eq:simmod}
\end{equation}
where $\bm{\mu}(\bm{\theta}^{(k)}, \bm{X}^{*}_i)$ and $\bm{\Sigma}_{\theta^{(k)}}$ are as defined in \eqref{eq:avgmodel}.  For example, assuming a (conservative) 80\% reduction in \ce{SO2} emissions occurs after FGD scrubber implementation, Figure~\ref{subfig:clifty} shows the forecasted mean decrease in \ce{SO4^2-} attributed to the Clifty Creek facility, calculated from 2000 simulated \ce{SO4^2-} surfaces as $\bm{\bar{V}}_{i} = \frac{1}{2000} \sum_{k = 1}^{2000} \hat{\bm{V}}^{(k)}_{i}$.

The utility of these forecasts is especially useful when assessing regulatory interventions.  Given possible resource limitations and the cost of installing FGD technologies, in what order should the five facilities be targeted for FGD implementation?  To address this question, we forecasted the population-averaged reduction in annual \ce{SO4^2-} exposure -- restricted to the population within the considered spatial domain -- after FGD installation at each facility (Figure~\ref{subfig:exp-est}), using draws from \eqref{eq:simmod}, an assumed 80\% reduction in \ce{SO2} emissions after scrubber installation, and gridded 2010 U.S. population density \citep{Falcone2016}. Figure~\ref{subfig:exp-est} suggests that adding scrubbers to the Kyger Creek facility should be a regulatory priority, as it results in the largest reduction in human exposure to \ce{SO4^2-} within the considered spatial domain. Importantly, despite emitting approximately \textit{twice} as much \ce{SO2} as the other power plants, the mean forecasted reduction in exposure due to FGD installation at Kyger Creek is about \textit{four times larger} than at the other facilities. The forecasted reduction in \ce{SO4^2-} exposure after intervention at the four other power plants appear largely the same; we thus recommend an intervention order determined by the FGD implementation cost at the remaining four facilities.  However, we are happy to report that, of the facilities still in operation in 2020 (Clifty Creek, Kyger Creek, Martin Lake, and Rockport), both Clifty Creek and Kyger Creek have FGD technologies in place to reduce \ce{SO2} emissions by 90\% \citep{Casey2020}.


\section{Discussion}
\label{sec:disc}

Models for spatial data are most often phenomenological, with a regression-based mean structure, and spatial autocorrelation modeled with a semiparametric random effect. This work proposes a new class of mechanistic models for spatial data that are constructed from OU processes. By first approximating the process with a discrete-space SDE, appropriate probability models can be constructed for spatial data viewed as (1) a temporal snapshot of the process, or (2) a time-averaged observation from the process. These models are versatile, and imply that inference on process parameters and probabilistic forecasts from transient and stationary space-time processes can be obtained from spatial data alone. As demonstrated in our analysis of 2011 \ce{SO4^2-} pollution and its relationship to sulfur dioxide emissions, these spatial models are especially useful for applications in which regulatory policy must be assessed.

It is important to compare the relative merits of our \ce{SO4^2-} pollution model against alternatives, including those more commonly found in atmospheric science. Our methodology is inspired by the rich history of mechanistic models for air pollution, including chemical transport models \citep{Stein2015}, plume models, and their so-called reduced-form hybrids \citep{Foley2014, Heo2016}. These models include detailed components for chemical transport, deposition, and reaction across 3-dimensional space and across time, and are typically used to assess point-source pollution via the simulation of many individual-level trajectories \citep{Henneman2019}. This level of detail and model complexity is especially powerful when simulating air pollution concentrations and evaluating pollution sources on a small geographic or temporal extent. However, the large computational burden and lack of data-driven inference limits the utility of such models when assessing aggregate data over large spatial and temporal regions \citep{Henneman2019}.  In contrast, our statistical model is directly tailored to \ce{SO4^2-} pollution aggregated over space and time, and we view its ability to infer (simplified) process dynamics and stochastic fluctuations on the time-scale of interest as its key advantage.

As shown in Section~\ref{subsec:humanexposure}, one advantage of our mechanistic model over existing statistical models for spatial data is its ability to provide probabilistic forecasts and uncertainty quantification for systems with varying initial conditions. This is especially useful when modeling systems directly impacted by human intervention, such as air pollution resulting from coal-fired power plant emissions. Because of the known relationship between sulfate pollution and human health, one interesting application of this methodology is as a model of spatial treatment interference (see \citet{ZiglerPapadogeorgou2018, Karwa2018}, and Section~\ref{sec:data}). For example, a causal analysis of the effect of FGD scrubber implementation on county-level human health outcomes requires a mapping of the treatment at power plant facility $i$ to all downwind population locations. Existing methodologies often seek to define such ``exposure mappings'' via static networks relating treatments to outcomes \citep{Karwa2018, Aronow2017}. However, given our knowledge of the physical system relating \ce{SO2} emissions to \ce{SO4^2-} concentrations and our model's ability to infer these dynamics, our mechanistic model may be a preferred specification of treatment exposure. We plan to explore the utility of this model as an exposure mapping, assessing how uncertainty and the presence of environmental confounders may be incorporated into existing causal inference methodology in the presence of interference.
 
Finally, the connections between the continuous space SPDE and discrete space SDE introduced in Section~\ref{subsec:STtoOU} presents a path for future development of mechanistic spatial models. Broader classes of SDEs, including alternative mean-reverting processes \citep{Allen2016} and models driven by L\'evy processes besides Brownian motion, may result in probability models for spatial data from a much broader class of space-time systems. Although we view discretization as an essential computational procedure for these models, additional connections between the implied spatial models constructed from the OU process and continuous space-time processes such as those presented by \citet{Brown2000} may prove fruitful in improving our understanding of the appropriateness of certain families of covariance functions for spatial data. 

\bigskip
\begin{center}
{\large\bf Appendix A: Derivation of the Time-Averaged Process}
\end{center}

The result follows from the well-known property \citep{Doob1942} that the integral of a Gaussian process $(Y_t)_{t \geq 0}$ with mean $\mu(t)$ and covariance $k(s, t)$ is itself a Gaussian process, where $V_T = \int_{0}^{T} Y_s ds$ has mean
\begin{equation}
\text{E}(V_T) = \int_{0}^{T} \mu(s) ds
\end{equation}
and variance
\begin{equation}
\text{Var}(V_T) = \int_{0}^{T} \int_{0}^{T} k(s, t) ds dt.
\end{equation}
Thus, for the space-time OU process defined by \eqref{eq:OU}, with $\bm{\mu}(t)$ and $k(s,t)$ given in \eqref{eq:GPmean} and \eqref{eq:GPcov}, we have
\begin{equation}
\int_{0}^{T} \bm{\mu}(s) ds = \int_{0}^{T} \bm{A}^{-1} \bm{m} ds = T \bm{A}^{-1} \bm{m},
\end{equation}
and
\begin{align}
\int_{0}^{T} \int_{0}^{T} & k(s, t) ds dt = \int_{0}^{T} \bigg( \int_{0}^{t} \bm{\Sigma} e^{-\bm{A}' (t - s)} ds + \int_{t}^{T} e^{-\bm{A} (s - t)} \bm{\Sigma} ds \bigg) dt \\
& = \int_{0}^{T} \bigg( \bm{\Sigma} (\bm{I} - e^{-\bm{A}' t}) (\bm{A}')^{-1} + (\bm{I} - e^{-\bm{A}(T - t)}) \bm{A}^{-1} \bm{\Sigma} \bigg) dt \\
& = T \big(\bm{\Sigma} (\bm{A}')^{-1} + \bm{A}^{-1} \bm{\Sigma} \big) - \bm{\Sigma} (\bm{I} - e^{-\bm{A}' T}) (\bm{A}')^{-2} - (\bm{I} - e^{-\bm{A}T}) \bm{A}^{-2} \bm{\Sigma}.
\end{align}
Finally, from the Lyapunov equation \eqref{eq:Lyap} we see that
\begin{equation}
\big(\bm{\Sigma} (\bm{A}')^{-1} + \bm{A}^{-1} \bm{\Sigma} \big) =  \big(\bm{A}' (\bm{B}\bm{B}')^{-1} \bm{A} \big)^{-1}
\end{equation} 
and the result follows.

\bigskip
\begin{center}
{\large\bf Appendix B: An Error Bound for the Time-Averaged Covariance}
\end{center}

Let $\bm{A} = (\gamma \bm{D} + \delta \bm{I})$ be an FDM or FVM approximation to a homogeneous diffusion process (rate of diffusion $\gamma$) with rate of decay $\delta$.  Then $\bm{A}$ is symmetric, and 
\begin{equation}
\bm{E} = \bm{\Psi} - \bm{\Phi} = \frac{-1}{T^2} (\bm{I} - e^{-\bm{A} T}) \bm{A}^{-3}.
\end{equation}
Under either periodic or zero flux boundary conditions, the diffusion operator matrix, $\bm{D}_{n \times n}$, is (1) symmetric with real elements, and (2) weakly diagonally dominant.  Thus, by the Gershgorin circle theorem, the eigenvalues $\lambda_i$ are all real and nonnegative, with $\lambda_{(n)} = \min_{i} \lambda_i = 0$.  Let $\bm{\Lambda}$ be a diagonal matrix with $\bm{\Lambda}_{ii} = \lambda_{i}$.  Factorizing $\bm{D}$ into its canonical form gives $\bm{D} = \bm{U} \bm{\Lambda} \bm{U}'$, where $\bm{U}$ is the square $n \times n$ matrix whose $i$th column is the eigenvector $u_i$ of $\bm{D}$.  Thus, $\bm{A}$ can be factorized as
\begin{equation}
\bm{A} = \gamma \bm{D} + \delta \bm{I} = \bm{U} \bm{V} \bm{U}',
\end{equation}
where $\bm{V}$ is a diagonal matrix with $\nu_i \equiv V_{ii} = \gamma \lambda_i + \delta$.  Note that $\nu_{(n)} = \gamma \lambda_{(n)} + \delta = \delta$.  Then,
\begin{align}
\vert \vert \bm{E} \vert \vert_{2} & = \vert \vert \frac{1}{T^2} (\bm{I} - e^{-\bm{A} T}) \bm{A}^{-3} \vert \vert_{2} \\
& =  \vert \vert \frac{1}{T^2} \bm{U} (\bm{I} - e^{-\bm{V} T}) \bm{V}^{-3} \bm{U}' \vert \vert_{2} \\
& \leq  \vert \vert \bm{U} \vert \vert_{2} \vert \vert \bm{U}' \vert \vert_{2} \vert \vert \frac{1}{T^2} (\bm{I} - e^{-\bm{V} T}) \bm{V}^{-3} \vert \vert_{2} \\
& = \vert \vert \frac{1}{T^2} (\bm{I} - e^{-\bm{V} T}) \bm{V}^{-3} \vert \vert_{2} \\
& = \frac{1}{T^2 \delta^3} (1 - e^{-\delta T}).
\end{align}
The final equality uses the fact that $\frac{1}{T^2} (\bm{I} - e^{-\bm{V} T}) \bm{V}^{-3}$ is a diagonal matrix with $i$th diagonal, $(1 - e^{-\nu_i T}) / (T^2 \nu_{i}^{3})$.  This function is decreasing in $\nu_i$.  Thus, $\vert \vert \frac{1}{T^2} (\bm{I} - e^{-\bm{V} T}) \bm{V}^{-3} \vert \vert_{2} = \frac{1}{T^2 \delta^3} (1 - e^{-\delta T})$.

\bibliography{ms}

\end{document}


\def\theequation{SM.\arabic{equation}}

\maketitle

\tableofcontents
\newpage

\section{Numerical Methods for Approximating Continuous Processes in Discrete Space}

Consider the continuous space-time stochastic process introduced in Section 3.1, 
\begin{equation}
dy_{\bm{s}}(t) = \bigg( - \mathcal{A}_{\bm{s}}(\bm{\theta}) y_{\bm{s}}(t) + m_{\bm{s}}(\bm{\theta}) \bigg) dt + \mathcal{B}_{\bm{s}}(\bm{\theta}) \; \xi(\bm{s}, t), \label{eq:cont-process}
\end{equation}
where $y_{\bm{s}}(t)$ ($\bm{s} \in \mathcal{D} \subset \mathbb{R}^2$, $t \geq 0$) is the quantity of interest, $\mathcal{A}_{\bm{s}}$ is a linear operator, $m_{\bm{s}}$ denotes  a source/sink at location $\bm{s}$, and $\mathcal{B}_{\bm{s}} \; \xi(\bm{s}, t)$ is a space-time white noise process, with process variance defined by real-valued function $\mathcal{B}_{\bm{s}}$. For convenience, we will assume that this process has constant variance, i.e., $\mathcal{B}_{\bm{s}} \equiv \sigma$ for all $\bm{s}$. As discussed in our manuscript, it is often useful to define a dynamic process as an SPDE in continuous space. However, the theoretical and computational limitations associated with working in continuous space-time necessitates an approximation of \eqref{eq:cont-process} in discrete space. This supplementary document provides a brief overview of how we implement such an approximation. Particular attention is focused on the sulfate model presented in the manuscript (see Sections 2 and 4 of the manuscript for details).

\subsection{Step One: Discretization}

The numerical representation of \eqref{eq:cont-process} in discrete space occurs in two steps. First, the continuous surface $\mathcal{D} \subset \mathbb{R}^2$ must be appropriately \textit{discretized} to some finite collection of points, $\mathcal{S} = \{\bm{s}_1, \dots, \bm{s}_k\}$. Second, the mathematical operator $\mathcal{A}$ is \textit{approximated with a matrix operator}, $\bm{A}$, on the set of discretized points, $\mathcal{S}$. The resulting stochastic process, defined on discrete space $\mathcal{S}$, is a computationally convenient representation of the original dynamic process defined in continuous space.

\begin{figure}[h]
\centering
\subcaptionbox{A $58 \times 35$ rectangular grid. \label{fig:grid-eg}}{\includegraphics[width=0.45\textwidth]{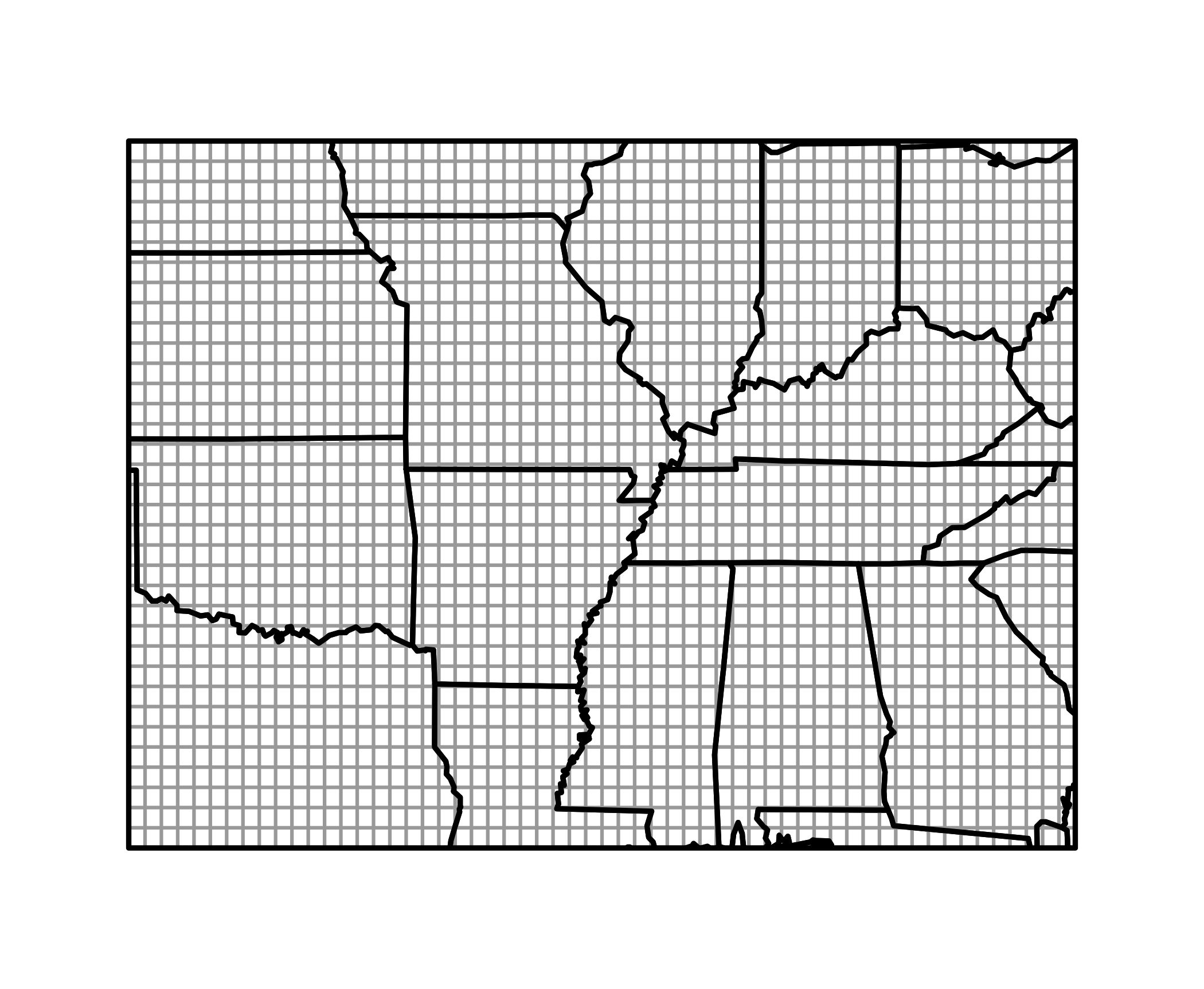}}
\hfill
\subcaptionbox{Mesh created with Delaunay triangulation. \label{fig:mesh-eg}}{\includegraphics[width=0.45\textwidth]{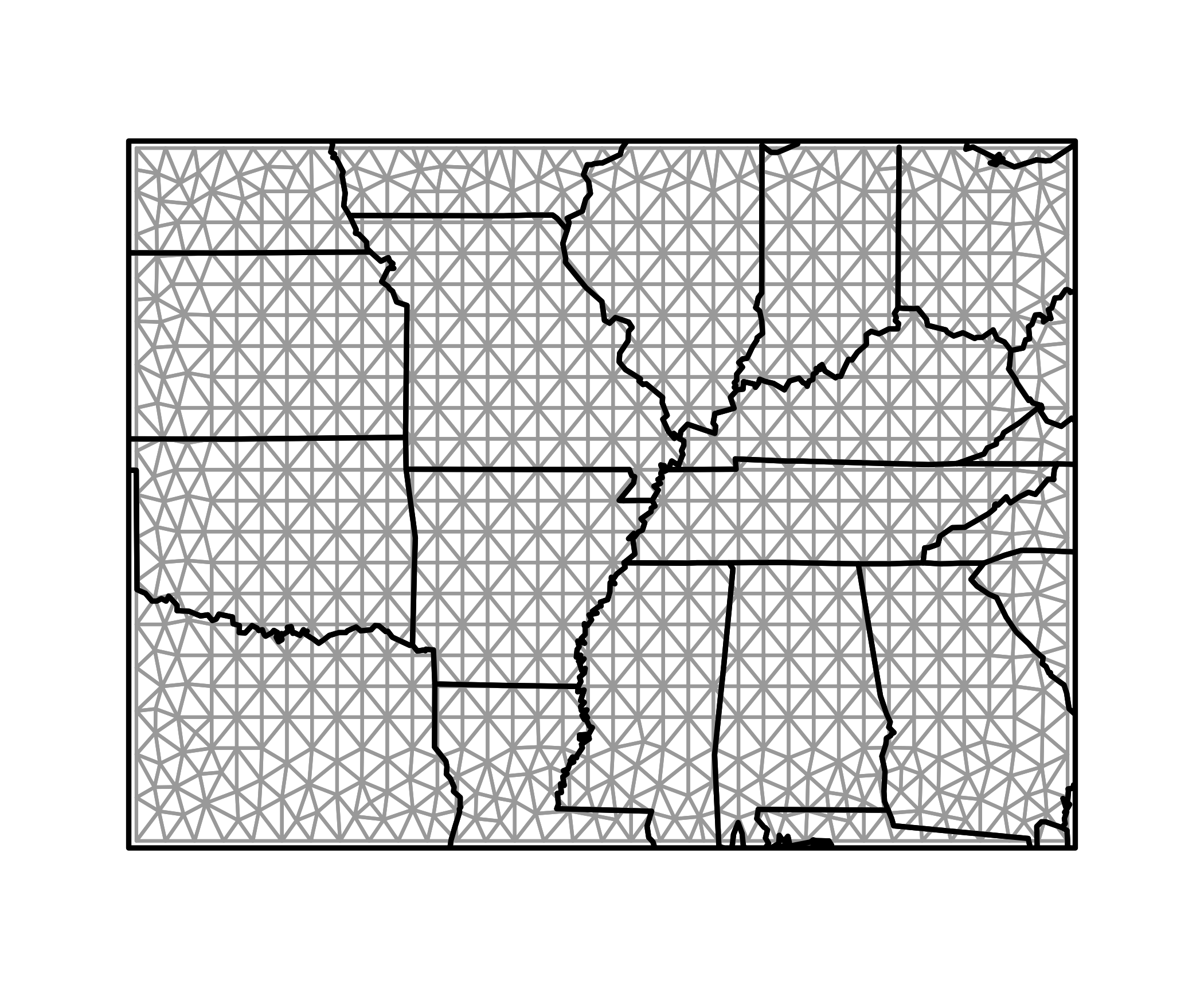}} 
\caption{Example discretizations of the surface, $\mathcal{D}$, considered in the sulfate analysis, including (a) a simple rectangular grid, and (b) a more complex Delaunay triangulation (created with the \texttt{R-INLA} package \cite{LindgrenRue2015}).}
\label{fig:disc-egs}
\end{figure}

The discretization step is conceptually simple. A grid or polygonal \textit{mesh} is overlaid on top of the continuous surface, $\mathcal{D}$, and the process dynamics are defined with respect to the discrete set of mesh cells. The structure of the mesh may vary, from a simple rectangular grid (Figure~\ref{fig:grid-eg}) to more complex structures, such as those generated via Delaunay triangulation (Figure~\ref{fig:mesh-eg}). The choice of mesh structure is often dictated by the complexity of the process dynamics, the boundary of the surface $\mathcal{D}$, and the chosen numerical approximation scheme (for example, finite element method implementations often use a triangulation method). Because of the variety of discretization schemes, and the large body of literature discussing their relative merits \cite{LangtangenLinge2017, VersteegMalalasekera2007, Johnson2009}, we encourage the interested reader to explore implementation details tailored to their problem of interest. For now, we emphasize that the discretization step restricts the process of interest to some discrete set of points, $\mathcal{S} = \{\bm{s}_1, \dots, \bm{s}_n\}$, corresponding to mesh vertices and/or centroids.

\subsection{Step Two: Numerical Approximation of $\mathcal{A}$}

After a suitable mesh has been chosen, process dynamics are approximated on the discretized space, $\mathcal{S}$. A variety of numerical schemes have been developed to facilitate this approximation, including the finite difference method (FDM), finite volume method (FVM), and finite element method (FEM).  Again, the literature and implementation for each of these methods are too extensive and problem-specific to permit a comprehensive introduction to these methods, and we encourage the interested reader to consult one of the following references for implementation details \cite{LangtangenLinge2017, VersteegMalalasekera2007, Johnson2009}. We instead emphasize that each method attempts to approximate a continuous mathematical operator, $\mathcal{A}$, with a discrete operator, $\bm{A}$, defined on the discretized space, $\mathcal{S}$. These approximations may occur via simple comparisons of neighboring values (such as first and second differences), or via the construction of appropriate basis functions; the implementation is dependent on the problem of interest and choice of numerical method.

As an example, consider the operator, 
\begin{equation}
\mathcal{A} = \gamma \Delta + \alpha \; \bm{v}_i(t) \cdot \nabla - \delta,
\end{equation}
corresponding to the advection-diffusion process used to model sulfate concentration in Section 2.1 (Equation (1) of the manuscript).  Here, $\gamma \Delta$ denotes homogeneous spatial diffusion with rate $\gamma$ (note, $\Delta \equiv \frac{\partial^2}{\partial x_{1}^2} +  \frac{\partial^2}{\partial x_{2}^2}$); advection due to wind is defined by the advective derivative $\bm{v}_i(t) \cdot \nabla \equiv v_{x_1} \frac{\partial}{\partial x_1} + v_{x_2}\frac{\partial}{\partial x_2}$, where $\bm{v}_i(t) = (v_{x_1}, v_{x_2})'$ is the velocity vector at time $t$; and atmospheric deposition occurs at rate $\delta$.  A simple numerical approximation of this operator might be constructed in the following way.

\begin{figure}[h]
\centering
\includegraphics[width=0.6\textwidth]{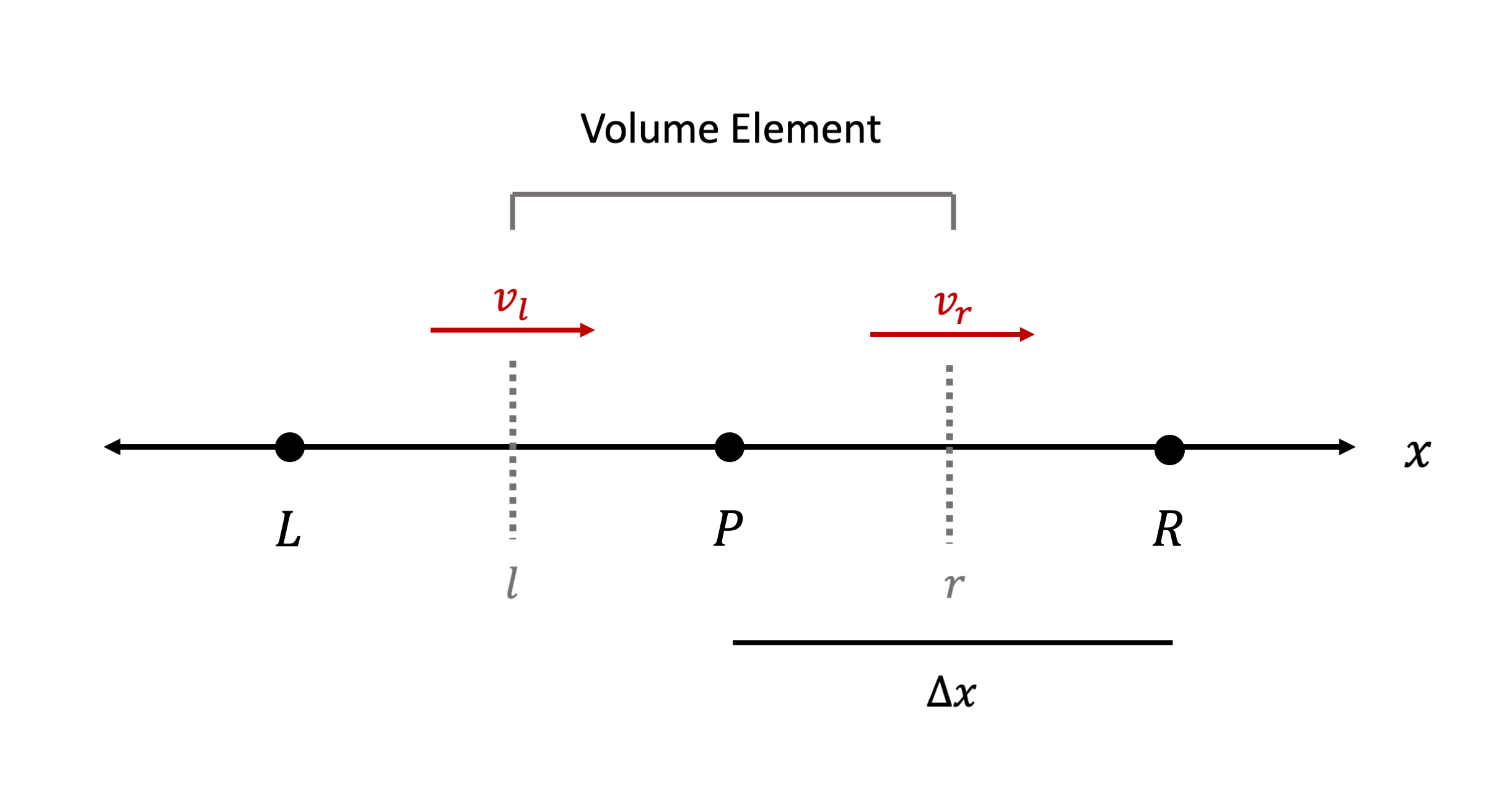}
\caption{Example discretization (regular grid, one dimension) at three points: $L$, $P$, and $R$. The process $Y(x,t)$ evaluated at these points is used for the second-order central difference approximation to diffusion and the FVM upwind scheme for advection-diffusion.}
\label{fig:points}
\end{figure}

Let $\mathcal{A}_1 \equiv \gamma \Delta$ denote homogeneous diffusion with rate $\gamma$, which for a process $Y(x, t)$ defined in one-dimensional space $x$ can be equivalently written as,
\begin{equation}
\mathcal{A}_1 \; Y = \frac{\partial}{\partial x} \bigg( \gamma \;  \frac{\partial Y}{\partial x} \bigg). \label{eq:diffusion}
\end{equation}
How might we approximate this operator on some set of regular discrete points, $\mathcal{S}$? Consider three points, $P$, $R = P + \Delta x$, and $L = P - \Delta x$, and assume the process $Y$ can be evaluated at each point at time $t$: i.e., $Y(P, t)$, $Y(R, t)$, and $Y(L, t)$. These correspond to three discrete points in our discretization, $\mathcal{S}$ (Figure~\ref{fig:points}). A second-order difference approximation of $\mathcal{A}_1$ can be constructed by first recognizing that the first derivative, $\frac{\partial Y(P, t)}{\partial x}$, can be approximated with simple linear differences,
\begin{equation}
\bigg[ \frac{\partial Y(P, t)}{\partial x} \bigg]_R \approx \frac{Y(P + \Delta x, t) - Y(P, t)}{\Delta x} =  \frac{Y(R, t) - Y(P, t)}{\Delta x} \label{eq:firstdiff1}
\end{equation}
and 
\begin{equation}
\bigg[ \frac{\partial Y(P, t)}{\partial x} \bigg]_L \approx \frac{Y(P , t) - Y(P - \Delta x, t)}{\Delta x} =  \frac{Y(P, t) - Y(L, t)}{\Delta x}. \label{eq:firstdiff2}
\end{equation}
Then, using \eqref{eq:firstdiff1} and \eqref{eq:firstdiff2}, we have
\begin{align}
\mathcal{A}_1 \; Y(P, t) & \equiv \frac{\partial}{\partial x} \big( \gamma \; \frac{\partial Y(P, t)}{\partial x} \big)  \\
 & \approx  \big( \gamma \; \big[ \frac{\partial Y(P, t)}{\partial x} \big]_R - \gamma \; \big[ \frac{\partial Y(P, t)}{\partial x} \big]_L  \big) / \Delta x \\
& = \gamma \; \frac{Y(R, t) - 2 Y(P, t) + Y(L, t)}{ (\Delta x)^2}.
\end{align}
Repeating this process for all points in $\mathcal{S}$, we can now approximate the operator $\mathcal{A}_1 \; Y$ in discrete space as,
\begin{equation}
\bm{A}_1 \; \bm{Y_t} =  \gamma \; \begin{pmatrix} 
    		\ddots & \ddots 	& \ddots 	& 		&  \\
   		  	  & 1 		& -2 		& 1 		& \\
    			  &        	& \ddots 	& \ddots 	& \ddots
    			\end{pmatrix}  \begin{pmatrix}
			\vdots \\
			Y(L, t) \\
			Y(P, t) \\
			Y(R, t) \\
			\vdots \end{pmatrix}.
\end{equation}
Note that $\bm{A}_1$ is a sparse, banded matrix, which permits computationally efficient matrix operations \cite{RueHeld2005} when working with the discrete space model.

The above discretization of a diffusion operator can be derived using either the finite difference method or the finite volume method \cite{VersteegMalalasekera2007}. However, the addition of advection due to wind necessitates a slightly more complex discretization scheme, as centered difference approximations become increasingly biased towards upstream process values as advective forces grow large \cite{VersteegMalalasekera2007}. Instead, we use an upwind differencing scheme to approximate advection-diffusion with the FVM. In short, for our three points, $P$, $R = P + \Delta x$, and $L = P - \Delta x$, we evaluate (or estimate) the wind velocity $v(x, t)$ along volume element faces $l$ and $r$ (i.e., $v_l \equiv v(l, t)$ and $v_r \equiv v(r, t)$; see Figure~\ref{fig:points}). Then, the operator
\begin{equation}
\mathcal{A}_2 \; Y \equiv  \gamma \Delta + \alpha \; \bm{v}_i(t) \cdot \nabla = \frac{\partial}{\partial x} \bigg( \gamma \;  \frac{\partial Y}{\partial x} \bigg) + \alpha \; v(x, t) \frac{\partial Y}{\partial x}
\end{equation}
can be approximated with 
\begin{equation}
\bm{A}_2 \; \bm{Y_t} =  \gamma \; \begin{pmatrix} 
    		\ddots & \ddots 	& \ddots 	& 		&  \\
   		  	  & a_L 	& -a_P 		& a_R 		& \\
    			  &        	& \ddots 	& \ddots 	& \ddots
    			\end{pmatrix}  \begin{pmatrix}
			\vdots \\
			Y(L, t) \\
			Y(P, t) \\
			Y(R, t) \\
			\vdots \end{pmatrix},
\end{equation}
where $a_L$, $a_P$, and $a_R$ are given in Table~\ref{tb:matrixentries}. Additional details of the FVM upwind differencing scheme can be found in Chapter 5 of Versteeg and Malalasekera \cite{VersteegMalalasekera2007}. 
\begin{table}[htb]
	\caption{Matrix entries for the FVM upwind difference approximation of an advection-diffusion operator.}
	\label{tb:matrixentries}
	\setlength{\tabcolsep}{12pt}
	\vspace{5mm}
	\renewcommand{\arraystretch}{1.8}
	\centering
	\resizebox{0.6\textwidth}{!}{
		\begin{tabular}{l r r r}
		 & \multicolumn{3}{c}{\textbf{Matrix Entries}} \\
		 \cline{2-4}
		\textbf{Edge Velocities} & $a_L$ & $a_R$ & $a_P$  \\
		\hline 
		$v_l > 0$, $v_r > 0$ & $\gamma + \alpha \; v_l$  & $\gamma$ & $2 \gamma + \alpha (v_r - v_l)$ \\
		$v_l < 0$, $v_r < 0$ & $\gamma$  & $\gamma - \alpha \; v_r$ & $2 \gamma + \alpha (v_r - v_l)$
	\end{tabular}
	}
\end{table}

Finally, including atmospheric deposition in the process (i.e., $\mathcal{A} = \gamma \Delta + \alpha \; \bm{v}_i(t) \cdot \nabla - \delta$) only requires the subtraction of the deposition rate $\delta$ from the matrix diagonal of $\bm{A}_2$, i.e., $\mathcal{A} \approx \bm{A} = \bm{A}_2 - \delta \bm{I}$. Thus, the advection-diffusion-deposition process found in our manuscript has been approximated in one dimension; the extension of these methods to two dimensions is straightforward \cite{VersteegMalalasekera2007}.  

\subsection{Atmospheric Sulfate Process}

\begin{figure}[htb]
\centering
\includegraphics[width=0.6\textwidth]{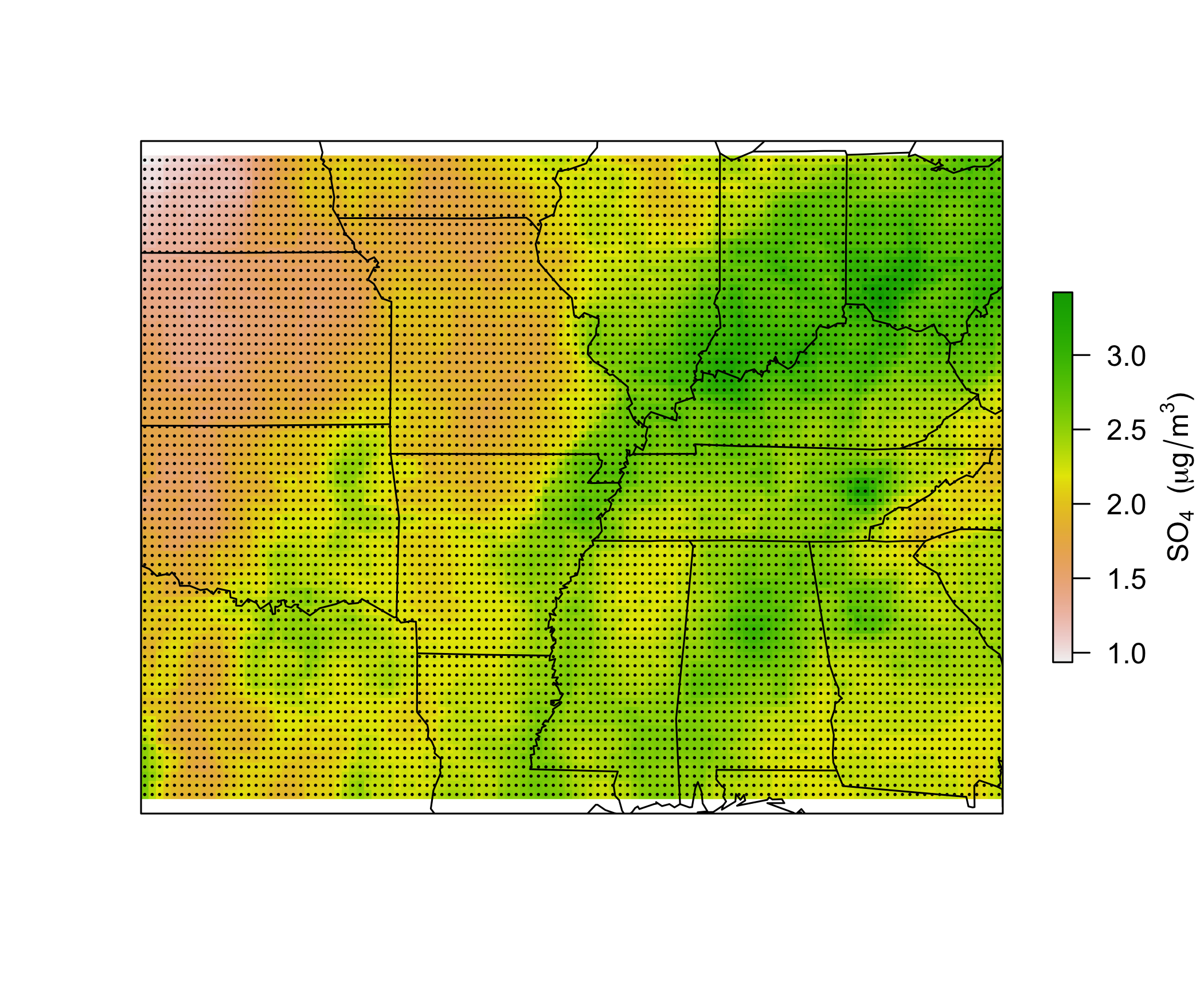}
\caption{Spatial region $\mathcal{D}$ considered in the sulfate analysis, with the overlaid $116 \times 70$ rectangular grid, $\mathcal{S}$, used in the discrete approximation.}
\label{fig:official-grid}
\end{figure}

In summary, the approximation of a continuous process in discrete space occurs by first discretizing continuous space $\mathcal{D}$ to discrete space $\mathcal{S}$, and then approximating the continuous operator $\mathcal{A}$ with its discrete representation, $\bm{A}$, as determined by some numerical scheme such as the FDM, FVM, or FEM. In our manuscript, the continuous process was the region of the continental USA shown in Figure~\ref{fig:official-grid}.  We discretized this space with a $116 \times 70$ rectangular grid, represented with black dots in Figure~\ref{fig:official-grid}. Then, using this discretization, we approximated the advection-diffusion-deposition process, $\mathcal{A} = \gamma \Delta + \alpha \; \bm{v}_i(t) \cdot \nabla - \delta$, with the $\mathbb{R}^2$ extension of the matrix $\bm{A}$ in the previous section. Finally, we let $\bm{Y_t}$ denote the vector of sulfate values evaluated at each point in $\mathcal{S}$ (Figure~\ref{fig:official-grid}), $\bm{m}$ a vector of sulfate sources corresponding to each point in $\mathcal{S}$, and $\bm{W_t}$ a vector of independent Brownian motions located at each point in $\mathcal{S}$. Thus, each continuous element in Equation~\eqref{eq:cont-process} has a discrete counterpart, and our continuous space-time sulfate process can now be approximated with the multivariate Ornstein-Uhlenbeck (discrete space) process,
\begin{equation}
d \bm{Y_t} =  \big(- \bm{A} \bm{Y_t} + \bm{m} \big) dt + \sigma \; d\bm{W_t}.
\end{equation}

\bibliography{supplement}
\bibliographystyle{plain}